\documentclass[prc,twocolumn,aps]{revtex4}

\usepackage{pstricks,pst-node,pst-text,pst-3d}
\usepackage{graphics}
\usepackage{epsfig}

\def\lsim{\mathrel{\rlap{\lower4pt\hbox{\hskip1pt$\sim$}}
    \raise1pt\hbox{$<$}}}         
\def\gsim{\mathrel {\rlap{\lower4pt\hbox{\hskip1pt$\sim$}}
    \raise1pt\hbox{$>$}}}         

\begin{document}
\title{Electron-deuteron scattering in the equal-time formalism:
beyond the impulse approximation}
\author{D.~R.~Phillips$^{a*,b}$, S.~J.~Wallace$^{c*,d}$,
  N.~K.~Devine$^{e}$ }
\affiliation{$^{a}$Department of Physics and Astronomy,
  Ohio University, Athens, OH 45701\\ 
  $^{b}$Department of Physics, University of Washington, Box 351560,
  Seattle, WA 98190\\
  $^{c}$Department of Physics, University of Maryland, College Park,
  Maryland 20742-4111 \\ 
  $^{d}$Theory Division, Jefferson Laboratory, 12000 Jefferson Avenue,
  Newport News, VA 23606\\
  $^{e}$General Sciences Corporation, 
  4600 Powder Mill Rd., Suite 400, Beltsville, Maryland 20705-1675}
\begin{abstract}
  Using a three-dimensional formalism that includes relativistic
  kinematics, the effects of negative-energy states, approximate
  boosts of the two-body system, and current conservation, we
  calculate the electromagnetic form factors of the deuteron up to
  $Q^2$ of 4 GeV$^2$.  This is done using a dynamical boost for
  two-body systems with spin. We first compute form factors in impulse
  approximation, but then also add an isoscalar meson-exchange current
  of pion range that involves the $\gamma \pi$ contact operator
  associated with pseudovector $\pi N$ coupling. We also consider
  effects of the $\rho \pi \gamma$ meson-exchange current.  The
  experimentally measured quantities $A$, $B$, and $t_{20}$ are
  calculated over the kinematic range probed in recent Jefferson
  Laboratory experiments. The $\rho\pi\gamma$ meson-exchange current
  provides significant strength in $A$ at large $Q^2$ and the
  $\gamma\pi$ contact-term exchange current shifts $t_{20}$, providing
  good agreement with the JLab data.  Relativistic effects and the
  $\gamma\pi$ meson exchange current do not provide an explanation of
  the $B$ observable, but the $\rho\pi\gamma$ current could help to
  provide agreement if a nonstandard value is used for the tensor
  $\rho N$ coupling that enters this contribution.
\end{abstract}
\maketitle 

\vspace{-0.8cm}

\section{Introduction}

The deuteron is a bound state of two spin-1/2 particles.
Consequently, the combined dictates of Poincar\'e invariance, current
conservation, and parity tell us that it has three independent form
factors. Conventionally these are taken to be the deuteron charge,
magnetic, and quadrupole form factors, $G_C$, $G_M$, and
$G_Q$, which are related to the Breit-frame matrix elements of the
deuteron electromagnetic current operator ${\cal A}_\mu$ in the three
deuteron magnetic sub-states $|+1 \rangle$, $|0 \rangle$, and $|-1
\rangle$ via the formulae:
\begin{eqnarray}
G_C&=&\frac{1}{3\sqrt{1 + \eta}e} (\langle 0|{\cal A}_0| 0 \rangle 
+ 2 \langle +1|{\cal A}_0|+1 \rangle), \label{eq:FC}\\
G_Q&=&\frac{1}{2 \eta \sqrt{1 + \eta} e} (\langle 0|{\cal A}_0| 0 \rangle
- \langle +1|{\cal A}_0|+1 \rangle),\\
G_M&=& \frac{-1}{\sqrt{2 \eta (1 + \eta)}e} \langle +1|{\cal A}_+|0\rangle.
\label{eq:FM}
\end{eqnarray}
(Here $\eta=Q^2/(4 M_d^2)$, and $Q^2=-q^2$ is the absolute value of
the square of the four-momentum transfer to the deuteron.) Three
experimental quantities are therefore required to disentangle the 
electromagnetic current of this nucleus. Two of these---the
structure functions $A$ and $B$---can be obtained from the
electron-deuteron differential cross-section using the usual
Rosenbluth separation:
\begin{equation}
\frac{d \sigma}{d \Omega}=\frac{d \sigma}{d \Omega}_{\rm Mott} \left[A(Q^2) 
+ B(Q^2) \tan^2\left(\frac{\theta_e}{2}\right)\right].
\end{equation} 
The structure functions $A$ and $B$ are related to $G_C$, $G_Q$,
and $G_M$, as follows:
\begin{eqnarray}
A&=&G_C^2 + \frac{8}{9} \eta^2 G_Q^2 + \frac{2}{3} \eta G_M^2,\\
B&=&\frac{4}{3} \eta(1 + \eta) G_M^2.
\end{eqnarray}
The third observable of choice is the dependence of the scattering on
the (tensor) polarization of deuterium. This can be measured either
using a polarized deuteron target (and unpolarized beam), or with an
unpolarized target by measuring the polarization of the recoil
deuterons. Both types of experiment result in the same tensor-polarization
observable, and in particular both can measure:
\begin{equation}
t_{20} \equiv -\sqrt{2} \, \frac{x(x+2)+ y/2}{1 + 2(x^2 + y)};
\end{equation}
where:
\begin{eqnarray}
x=\frac{2 \eta G_Q}{3 G_C}; \, \,
y=\frac{2 \eta}{3} \left[\frac{1}{2} + (1 + \eta)
\tan^2\left(\frac{\theta_e}{2}\right)\right] \frac{G_M^2}{G_C^2}.
\end{eqnarray}

Recent experiments at the Thomas Jefferson National Accelerator
Facility (JLab) have probed the electromagnetic form factors of the
deuteron at large space-like momentum transfers. $t_{20}$ has been
measured at $Q^2$ up to almost 2 GeV$^2$~\cite{Ab00A,Ab00B}, $B$ out
to about 1.3 GeV$^2$, and $A$ to $Q^2=6$ GeV$^2$~\cite{Al99,Ab99}.
Two recent reviews provide up-to-date information on these experiments
and their theoretical interpretation~\cite{GarconOrden01,GilmanGross02}.

The kinematic range of these data pushes the limit of theoretical
descriptions of this simplest of nuclei. At small values of $Q^2$ the
appropriate degrees of freedom for this description are
nonrelativistic nucleons, interacting via static nucleon-nucleon
interactions, with small corrections due to meson-exchange currents
and relativistic effects~\cite{RS91,Wi95,Schiavilla02,Ar99,Si01}. Provided
$Q^2$ is below the scale of chiral symmetry breaking, contributions to
electron-deuteron scattering can be organized using the nuclear
effective theory first proposed by Weinberg~\cite{We90,We91,We92}, and
the results are in good agreement with the extant data~\cite{Ph03}
below $Q^2 \sim 0.5$ GeV$^2$. As $Q^2$ increases to $\sim 1$ GeV$^2$,
nucleon and meson degrees of freedom may still be appropriate but a
relativistic formalism is needed to account for relativistic
kinematics and dynamics.  At very high values of $Q^2$ one expects a
transition to a regime where quark and gluon degrees of freedom
provide the most natural description of data.  To date, the
experimental and theoretical situation for elastic electron-deuteron
scattering can be summarized by saying that there is no significant
evidence of having reached the quark regime~\cite{GarconOrden01}. The
most important aspects of the internal quark structure of a nucleon
appear to be taken into account via nucleon form factors and
relativistic $NN$ dynamics.

Considerable effort has been invested in the construction of these
relativistic formalisms for the $NN$ bound state and associated
meson-exchange currents. The goal is to address elastic
electron-deuteron scattering at $Q^2$ of a few GeV$^2$, see
e.g.~Refs.~\cite{PW96,PW97,PW98,HT,vO95,Ar80,Ch88,CK99}.  Such an
approach is a logical extension of the standard nonrelativistic
treatment of the $NN$ system (see, e.g.~Ref.~\cite{CS98}) which has
had a significant amount of success in describing $A$, $B$, and
$t_{20}$ using a non-relativistic $NN$ interaction that is fit to the
$NN$ scattering data (see,
e.g.~Refs.~\cite{RS91,Wi95,Schiavilla02}). Like the non-relativistic
approach, relativistic approaches to the $NN$ problem are grounded in
a phenomenological description of the $NN$ scattering data. However,
they seek to implement dynamics that obeys the Poincar\'e
algebra---even if only approximately---and in so doing they go beyond
the non-relativistic treatment of the $NN$ system. If electromagnetic
interactions with the deuteron are also to be considered
then---regardless of the momentum transfer involved---it is crucial
that the consequences of electromagnetic gauge invariance be
incorporated in the calculation.  Minimally this means that the
electromagnetic current of the deuteron must be conserved.  Indeed,
the derivation of Eqs.~(\ref{eq:FC})-(\ref{eq:FM}) {\it assumed} that
the deuteronic current ${\cal A}_\mu$ was conserved.  This motivates
the use of meson-exchange interactions for which methods of quantum
field theory may be used to construct the conserved current.

The three-dimensional ``equal-time'' (ET) formalism which was
developed and applied in Refs.~\cite{PW96,PW97,PW98} is one such
method.  This approach starts from the four-dimensional Bethe-Salpeter
formalism and the Mandelstam construction of the electromagnetic
current.  It includes relativistic kinematics, negative-energy states
and relativistic pieces of the electromagnetic current explicitly at
all stages of the calculation.  In Ref.~\cite{PW98} we reported on
impulse-approximation calculations of the form factors
(\ref{eq:FC})--(\ref{eq:FM}).  The Bonn-B interaction~\cite{Ma89},
that was fit to $NN$ scattering data using a relativistic equation
with only positive-energy states, was used.  In contrast to
Ref.~\cite{Ma89} we employed pseudovector $\pi N$ coupling, since
it is then easier to implement the constraints of chiral
symmetry. 
In Ref.~\cite{PW98} we focused particularly on the role of
negative-energy components of the deuteron wave function in the ET
formalism.  The $NN$ interaction with negative-energy components was
obtained by adjusting the $\sigma NN$ coupling so that the 
deuteron binding energy was the same as in Ref.~\cite{Ma89}. Once this
was done the inclusion of negative-energy components of spinors
produced only modest effects: there were noticeable changes to some
deuteron observables at larger momentum transfers but not much
improvement in the description of the experimental data.  This is in
agreement with earlier results of Hummel and Tjon~\cite{HT}, and is
not surprising since pseudovector coupling suppresses the coupling to
the negative-energy states.

The results of Ref.~\cite{PW98} for observable $t_{20}$ were in good
agreement with experiments but the description of the deuteron
magnetic form factor was poor.  Similar conclusions held
regardless of whether or not the negative-energy components of the
deuteron wave function were included.  Some other relativistic
approaches, e.g.~Ref.~\cite{vO95}, have found larger effects
for negative-energy components and this point is discussed further in 
Sec.~\ref{sec-results}.
  
However, a number of important effects were not included in the
impulse-approximation calculation of Ref.~\cite{PW98} (hereafter
``ETIA''). These included the dynamical boost of bound-state vertex
functions, an isoscalar pionic meson-exchange contribution to ${\cal
A}_\mu$ that arises from use of pseudovector $\pi N$ coupling, and the
$\rho\pi\gamma$ meson-exchange current. In the present work we use the
positive-energy-state ETIA calculation as a baseline. We then include
the various effects listed above that were omitted in
Ref.~\cite{PW98}, and also display the contribution of Z-graphs
computed to first order in perturbation theory.

These different contributions to ${\cal A}_\mu$ can be organized
according to the power counting developed by
Weinberg~\cite{We90,We91,We92} that has already been successfully
applied to a number of electromagnetic reactions involving deuterium.
Since the power counting is based on an expansion in powers of
\begin{equation}
{\cal P} \equiv \frac{|\vec{p}|,m_\pi,|\vec{q}|}{\Lambda},
\end{equation}
with $\Lambda$ the scale of chiral symmetry breaking, it breaks down at
$Q^2 \sim 0.5$ GeV$^2$, but examining the different contributions to
${\cal A}_\mu$ using the nuclear effective theory provides a way to
anchor their low-momentum behavior in a systematic way.

In this counting the calculation of Ref.~\cite{PW98} included all
mechanisms of $O(e)$ and $O(e{\cal P})$. Formally the most important
neglected mechanism occurs at $O(e{\cal P}^2)$, and arises because
electromagnetic currents are evaluated in the Breit frame, where the
deuteron has momentum $-{1 \over 2}{\bf q}$ in the initial state and
momentum $+{1 \over 2}{\bf q}$ in the final state. In this case the
deuteron wave functions must be boosted from the deuteron rest frame
to obtain the initial and final-state wave functions in the Breit
frame. The ET formalism developed in Ref.~\cite{PW96,PW97,PW98} was
not boost invariant, and this failure to respect Poincar\'e invariance
resulted in an error at $O(e{\cal P}^2)$ in the calculation of
Ref.~\cite{PW98}.  In order to remedy this, a boost rule for scalar
particles was developed in Ref.~\cite{Wa01} where it was shown to
provide an exact result for the boost of the two-body energy.  In this
work, the boost rule is extended approximately to the case of two
interacting spin-1/2 particles as discussed in Sec.~\ref{sec-boost}.
We have verified numerically that the ET equation together with the
boost rule for the $NN$ interaction provides eigenstates of the
deuteron with energy $E_{\bf P} = \sqrt{M^2 + {\bf P}^2}$ when the
total momentum is ${\bf P}$. The deuteron mass $M$ is therefore
invariant for the total momentum values of interest in this work.

An exact solution of the boost problem has been given in the instant
form of dynamics by Coester and Polyzou~\cite{CP82}, following the
methods suggested by Bakamjian and Thomas~\cite{BT53}. In that
solution, interactions are introduced solely into the rest-frame mass
operator. In contrast, here we use the ET reduction of field-theoretic
expressions to find an effective hamiltonian as well as currents that
are consistent with that hamiltonian. (In particular, we ensure that
the currents satisfy a Ward-Takahashi identity.) In the two-body
rest-frame system, the hamiltonian found via our ET reduction could be
used to define a suitable mass operator, which could then be used to
implement an exact boost to other frames following the methods
of Ref.~\cite{CP82}. We do not employ this ``exact boost''
construction because the electromagnetic currents that are obtained
from the ET reduction are not consistent with it.  Instead, in our
work we develop an approximate boost rule which maintains a clear
correspondence to field theory, and so facilitates a more consistent
treatment of strong and electromagnetic interactions.

Another important contribution not included in Ref.~\cite{PW98} enters
at $O(e{\cal P}^3)$. When pseudovector (PV) $\pi N$ coupling is
employed an additional contribution to the two-body isoscalar charge
operator---first identified by Riska~\cite{Ri84}---must be
included. Without it the unitary equivalence between PV and
pseudoscalar $\pi N$ coupling will not be respected~\cite{Friar}. In
this work we include this PV-coupling current and find that it is
significant, albeit not as important as when it is added in some
nonrelativistic approaches~\cite{RS91}.  The decreased importance of
this effect in our approach is traceable to a factor of half in the
PV-coupling current in the ET formalism relative to the current used
in Ref.~\cite{RS91}. This factor must be taken into account if the
unitary equivalence between different techniques for obtaining
relativistic corrections to the $NN$ interaction and currents is to be
maintained~\cite{Ad93}.

The $\rho\pi\gamma$ meson-exchange current (MEC) has often been
invoked in calculations of electron-deuteron scattering. In this MEC
the photon interacts with the $\rho$-meson cloud of one nucleon,
producing a pion which is absorbed by the other nucleon.  Thus the
$\gamma N$ interaction is of short range but the MEC has long range
because a pion exchange is involved.  This MEC contributes to $G_M$ at
$O(e{\cal P}^4)$ in the nuclear effective theory, but as
we show it can produce substantial
effects in the magnetic form factor at large $Q$. We include it in our
relativistic calculation and consider the extent to which it can help
to explain the magnetic form factor in the ET formalism.  Although
there is not much improvement in the description of data if the
standard tensor $\rho N$ coupling of one-boson-exchange $NN$ models is
adopted, other values of $f_\rho/g_\rho$ may help to explain
$B(Q^2)$. Note that we do {\it not} include the $\omega\sigma\gamma$
current that has been considered by Hummel and Tjon \cite{HT} because
it is of shorter range.

Our paper is structured as follows. First we give a brief review of
the ET formalism in which we display expressions for the bound-state
equation, the $NN$ interaction, and the current matrix element---all
for the particular case of an instantaneous two-body interaction.
Second, we discuss the boost of the bound state wave function that is
needed to evaluate matrix elements in the Breit frame. Third, we
discuss inclusion of meson-exchange-current contributions, especially
the PV-coupling current and the $\rho \pi \gamma$ MEC, each of which
has been found to give significant contributions to electron-deuteron
scattering.  We then also describe the inclusion of Z-graph effects
(coupling to the negative-energy states) in perturbation
theory. Lastly, we present our results for $A$, $B$, and $t_{20}$ when
each of these effects is added to the baseline ETIA calculation.

\section{The equal-time approach}

A number of alternative 3D relativistic treatments of deuteron
dynamics exist (see, for instance,
Refs.~\cite{HT,vO95,Ar80,Ch88,CK99}). Of these, the formalism that is
closest to this work is that of Hummel and Tjon~\cite{HT}, although
here we eliminate some approximations made in Ref.~\cite{HT}.  In this
section we summarize the development of the ET formalism with respect
to obtaining the one-body limit and a systematic reduction from 4D to
3D.

Consider the 4D Bethe-Salpeter equation (BSE) for a bound-state vertex
function, $\Gamma$:

\begin{equation}
\Gamma=K G_0 \Gamma.
\label{eq:LBSE}
\end{equation}
Here $K$ is, in principle, the sum of all two-particle-irreducible $NN
\rightarrow NN$ graphs. The $NN$ propagator $G_0$ is the product of
spin-half Feynman propagators for each nucleon: $G_0=i d_1 d_2$.  In
studies of this equation for the deuteron bound state~\cite{FT} the
kernel $K$ included a set of single-boson exchanges---in analogy to
many non-relativistic potential models---yielding the ``ladder''
approximation.  However, it is well known that in such an
approximation the Bethe-Salpeter equation does not give the correct
one-body limit~\cite{Gr82}. In other words, if we consider
unequal-mass particles, and take one of them to be very heavy,
Eq.~(\ref{eq:LBSE}) does not become the Dirac equation for the light
particle moving in the static field of the heavy one. This limit is
only properly treated in Eq.~(\ref{eq:LBSE}) if the full set of ladder
and crossed-ladder graphs is taken for $K$~\cite{Gr82}. In
Ref.~\cite{PW96} we provided a remedy to this flaw, and showed that the
pieces of the graphs that appear in $K$ and are responsible for the
one-body limit can be resummed so that Eq.~(\ref{eq:LBSE}) becomes:
\begin{equation}
\Gamma=U (G_0 + G_C) \Gamma,
\label{eq:4DET}
\end{equation}
where the precise form of $G_C$ was derived in \cite{PW96,PW98}.  For
exact correspondence between (\ref{eq:LBSE}) and (\ref{eq:4DET}) we
should have:
\begin{equation}
K=U + U G_C K.
\end{equation}
At the level of the one-boson-exchange interaction, where $K$ and $U$
have only their lowest-order pieces, we see that Eq.~(\ref{eq:4DET})
defines an improved ``ladder'' Bethe-Salpeter equation, which {\it does}
have the correct one-body limit:
\begin{equation}
\Gamma=K^{(2)} G \Gamma,
\label{eq:LBSEplus}
\end{equation}
where $G = G_0 + G_C$. Note that if the field theory to  be solved
involves nucleons and mesons then the approximation $K \rightarrow
K^{(2)}$ is equivalent to restricting ourselves to a one-boson-exchange
kernel.

This equation is still four-dimensional.  The next step is to perform
a systematic reduction to three dimensions.  We can motivate the
reduction scheme by following Salpeter~\cite{Sa52} and assuming that
the dominant interaction is the instantaneous part, i.e., the part
obtained by the replacement:
\begin{equation}
K^{(2)}(q)=\frac{\Gamma_1 \Gamma_2}{q^2 - \mu^2} \quad \longrightarrow \quad 
K^{(2)}_{\rm inst}({\bf q})=-\frac{\Gamma_1\Gamma_2}{{\bf q}^2 + \mu^2},
\label{eq:replace}
\end{equation}
where $\Gamma_1$ and $\Gamma_2$ are Dirac-matrix structures of the
meson-nucleon vertices and $q=(q_0,{\bf q})$ is the four-momentum of
the meson.  Since $K^{(2)}_{\rm inst}$ depends only on the
three-vector $\bf q$ this replacement reduces Eq.~(\ref{eq:LBSEplus})
to a three-dimensional equation.  If we were to stop at this point,
the reduction would not be systematic nor would it include retardation
effects.  A central result of Ref.~\cite{PW96} was that the
three-dimensional reduction could be implemented in a systematically
improvable way.  Instead of simply neglecting the retardation effects,
they can be reorganized into a 3D interaction kernel $K_{\rm ET}$.
This interaction shares the property of $K^{(2)}_{\rm inst}$ that it
does not depend upon the time-component of momentum transfer.  If the
reduction includes only positive-energy matrix elements, then $K_{\rm
ET}$ is the 3D interaction obtained from time-ordered perturbation
theory.

Thus the reduction to three dimensions produces the following
two-body equation: 
\begin{equation}
\Gamma_{\rm ET}=K_{\rm ET} \langle G \rangle \Gamma_{\rm ET}.
\label{eq:ET}
\end{equation}
Here the three-dimensional propagator $\langle G \rangle$ is obtained 
by integrating over the time-component of relative four-momentum:
\begin{equation}
\langle G \rangle \equiv \int \frac{dp_0}{2 \pi}
\left[ G_0(p;P) + G_C(p;P) \right]
\end{equation} 
and the ET interaction is defined in lowest order by
\begin{equation}
\langle G \rangle K_{\rm ET} \langle G \rangle \equiv
\langle G K G \rangle.
\label{eq:ETK}
\end{equation}
(Hereafter we always denote integration over zeroth components of
relative four-momenta by angled brackets.)

In order to explain the significance of the $G_C$ term in $\langle G
\rangle$ and its connection to Z graphs and the one-body limit,
consider the more standard equal-time Green's
function~\cite{LT63,BK93B} that omits $G_C$.  It is sufficient to work
in the c.m. frame of two particles of equal masses because the boost
discussed in Sec.~\ref{sec-boost} will provide the results needed in
other frames.  Then one finds
\begin{equation}
\langle G_0 \rangle=\frac{\Lambda_1^+ \Lambda_2^+}{E - 2 \epsilon
} - \frac{\Lambda_1^- \Lambda_2^-}{E +  2 \epsilon},
\label{eq:aveG0}
\end{equation}
where $\Lambda^{\pm}$ are related to projection operators onto
positive and negative-energy states of the Dirac equation, $E$ is the
total energy, and $\epsilon=({\bf p}^2 + m^2)^{1/2}$. The
propagator $\langle G_0 \rangle$ is not invertible~\cite{Kl}, since it
has no components in the $+-$ and $-+$ sectors. This is related to the
lack of a correct one-body limit in the ladder BSE.  If we had applied
the 3D reduction (\ref{eq:replace}) to Eq.~(\ref{eq:LBSE}) we would
have obtained the Salpeter equation:
\begin{equation}
\Gamma_{\rm S}=K_{\rm inst} \langle G_0 \rangle \Gamma_{\rm S},
\label{eq:Salpeter}
\end{equation} 
which has the non-invertible $\langle G_0 \rangle$ in the intermediate
state.  However, adding $G_C$---which comes from
resumming pieces of the crossed-ladder graphs---{\it before} reducing to
three dimensions gives a 3D $NN$ propagator:
\begin{eqnarray}
  \langle G \rangle&=&\frac{ \Lambda_1^+ \Lambda_2^+}{E -
    2 \epsilon } - \frac{ \Lambda_1^+ \Lambda_2^-}{
    2 \epsilon } -
  \frac{ \Lambda_1^- \Lambda_2^+}{ 2 \epsilon
    } - \frac{ \Lambda_1^- \Lambda_2^-}{E + 2 \epsilon }.
\label{eq:aveG0GCgeneral}
\end{eqnarray}
This is the three-dimensional propagator that was derived by Mandelzweig
and Wallace~\cite{MW}, here specialized to the c.m. frame for two
equal-mass particles.  In its more general form for
unequal masses, the contribution $\langle G_C \rangle$
to $\langle G \rangle$ provides the
correct one-body limit as {\it either} particle's mass
tends to infinity.  The propagator also is invertible.

\begin{figure}[h,t,b]
\includegraphics[width=2in]{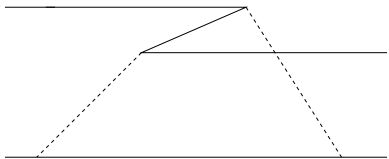}

\vskip 3 mm

\caption{One example of a Z-graph which is included in our
3D equation (\ref{eq:ET}).
\label{fig-Zgraph}  }  
\end{figure}

With regard to Z-graphs, one may compare the $++ \rightarrow ++$ piece of 
\begin{equation}
K^{(2)}_{\rm ET} \langle G \rangle K^{(2)}_{\rm ET}
\end{equation}
with the amplitude obtained at fourth order in the full 4D field
theory.  We find that the contribution of negative-energy states
agrees at leading order in $1/m$~\cite{PW98}, and that this would be
true even were different mass particles considered and the mass
of either one taken to infinity. In other words, effects such as
Fig.~\ref{fig-Zgraph} are included in a bound-state calculation that
employs Eq.~(\ref{eq:ET}). And this is true even if only the instantaneous
ladder kernel $K_{\rm inst}^{(2)}$ is used, because of our careful
treatment of the one-body limit.


\section{The instantaneous $NN$ interaction}

In Ref.~\cite{PW98} we examined the importance of retardation effects
in $ K_{\rm ET}$. We found that they had little impact on
deuteron electromagnetic form factors. Hence in this work we report
only on results obtained using Eq.~(\ref{eq:ET}) with an instantaneous
interaction.  This is consistent with use of the Bonn-B interaction
that takes an instantaneous form in the center-of-mass frame.  Other
reductions to three dimensions are certainly possible but the result
that retardation effects are small should not be sensitive to the
reduction used.

As an example of how the reduction to an instantaneous,
three-dimensional interaction is performed we consider the
four-dimensional pseudovector pion kernel:
\begin{equation}
K^{\pi} = \frac{g_{\pi}^2\tau_1\cdot \tau_2}{q^2 - m_{\pi}^2} 
\frac{\gamma_1^5 \gamma_1 \cdot (p_1 - p'_1)}{2m}
\frac{\gamma_2^5 \gamma_2 \cdot (p_2 - p'_2)}{2m}.
\label{eq:4DKpi}
\end{equation}
We use the identity:
\begin{eqnarray}
&&\bar{u}_1' \gamma^5_1 \frac{\gamma_1 \cdot (p_1 - p_1')}{2m}u_1 = 
\nonumber \\
&& \qquad   \bar{u}_1' \Big\{ \gamma^5_1  
+ \frac{(p_1^0 - {p_1'}^0 -\epsilon_1 + \epsilon_1')}{2m}
 \gamma_1^5 \gamma_1^0 \Big\} u_1,
\end{eqnarray}
where $u_1'$ and $u_1$ are Dirac spinors for positive-energy states,
and carry out integrations over time-components of momenta as implied
by the brackets in Eq.~(\ref{eq:ETK}).  To derive the
instantaneous interaction we then take the static limit, i.e. assume
$|E - \epsilon -\epsilon'| \ll \omega$. In the language of effective
field theory this means we consider only the effects of ``potential
pions'' (which have $q_0 \sim 0$) and ignore the impact of ``radiation
pions'' ($q_0 \sim m_\pi$) on our results~\cite{Ka98}. The result for
the instantaneous ET interaction in $++$ states is then:
\begin{eqnarray} 
&& K^{\pi}_{\rm ET} = - \frac{g_{\pi}^2}{{\bf q}^2 + m_{\pi}^2} 
\bar{u}_1' \bar{u}_2' \Big\{ \gamma_1^5 \gamma_2^5 
-  \frac{d_1 d_1'}{4m^2} \gamma_1^5 \gamma_1^0 \gamma_2^5 \gamma_2^0
\nonumber\\
&& \qquad + \frac{d_1 - d_1'}{4m}
\left( \gamma_1^5 \gamma_1^0 \gamma_2^5  +  
\gamma_1^5 \gamma_2^5 \gamma_2^0 \right) \Big\} u_1 u_2,
\label{eq:KpiET}
\end{eqnarray}
where $u_1$ and $u_2$ are Dirac spinors, $d_1 = E_{\bf P} -
\epsilon_1({\bf p}_1) - \epsilon_2({\bf p}_2)$ and $d_1' = E_{\bf P} -
\epsilon_1({\bf p}'_1) - \epsilon_2({\bf p}'_2)$.  This interaction
differs from a pseudoscalar one because of the terms involving $d_1$
and $d_1'$, which are sub-leading in $1/m$.  They arise
because---although we ignored retardation in the denominator of
Eq.~(\ref{eq:4DKpi})---we must still choose an energy shell to
determine the value of $p_1^0 - {p_1'}^0$.  The interaction
(\ref{eq:KpiET}) agrees with the $\tilde{\mu}=1$ interaction used by
Adam, G\"{o}ller, and Arenh\"{o}vel~\cite{Ad93}, up to terms of order
$p^4/m^4$ . 
 
\section{Current conservation in the equal-time approach}
\label{sec:current-conservation}

In order to compute deuteron electromagnetic form factors we must also
construct a conserved electromagnetic deuteron current.  In four
dimensions there are two pieces to the impulse current, the first
being determined by the $G_0$ part of the propagator:
\begin{eqnarray}
&& \bar{\Gamma}(P+q) G_0' J_{0,\mu}G_0 \Gamma(P) =\nonumber\\ 
&& \quad i  \bar{\Gamma}(P+q)  d_1(p_1) \, d_2(p_2 + q) 
 j_\mu^{(2)} d_2(p_2) \Gamma(P)   + (1 \leftrightarrow 2),\nonumber\\
&& 
\label{eq:BSEIA}
\end{eqnarray}
where $\Gamma$ is the solution of Eq.~(\ref{eq:LBSE}) for initial deuteron 
momentum $P= p_1 + p_2$, $\Gamma'$ is the solution for final deuteron momentum $P+q$,
$d_n$ is the Dirac propagator for particle $n$, and $j_\mu^{(n)}$
is the usual one-nucleon current for particle $n$:  

\begin{equation}
  j_\mu^{(n)}=e^{(n)} \left[F_1^{(n)}(Q^2) \gamma^{(n)}_\mu + F_2^{(n)}(Q^2)
    \frac{i}{2 m} \sigma_{\mu \nu}^{(n)} q^\nu \right] 
\label{eq:job}
\end{equation}
where $e^{(n)}=\frac{|e|}{2}(1 + \tau_3^{(n)})$ is the charge.  Using
the Ward-Takahashi identity associated with formally modified but
practically identical form of this current~\cite{GR87} it is easy to
show that the current (\ref{eq:BSEIA}) is conserved, i.e.
\begin{equation}
q^\mu \bar{\Gamma}(P+q) G_0' J_{0,\mu}G_0 \Gamma(P) =0,
\end{equation}
provided that $\Gamma$ obeys the ladder Bethe-Salpeter equation.

To obtain an equivalent three-dimensional current, we can replace
$\Gamma \rightarrow \Gamma_{\rm S}$, yielding
the conserved current
\begin{eqnarray}
&& {\cal A}_{\rm S,\mu}=\bar{\Gamma}'_{\rm S} 
\left\{ \langle G_0' J_\mu G_0 K^{(2)} G_0 \rangle
 + \langle G_0' K^{(2) \prime} G_0' J_\mu G_0 \rangle\right.\nonumber\\
&& \qquad \qquad \qquad \qquad \qquad \qquad \qquad
\left.- \langle G_0' J_\mu^{} G_0 \rangle \right\} \Gamma_{\rm S},
\label{eq:3termcurrent}
\end{eqnarray}
where $\Gamma'$, $G_0'$ and $K^{(2) \prime}$ include the momentum $Q$ from the
photon absorption. Since we are considering the equal-time case we can
also replace $K^{(2)} \rightarrow K_{\rm inst}$, and because $K_{\rm inst}$ has
no dependence on time components of momenta, that replacement applied
to Eq.~(\ref{eq:3termcurrent}) leads to
\begin{eqnarray}
{\cal A}_{\rm S,\mu}=\bar{\Gamma}'_{\rm S} \{ \langle G_0' J_{\mu} G_0
\rangle K_{\rm inst} \langle G_0 \rangle &+& \langle G_0' \rangle
K_{\rm inst} \langle G_0' J_{\mu} G_0 \rangle \nonumber\\ &-& \langle
G_0' J_{\mu} G_0 \rangle\} \Gamma_{\rm S}
\label{eq:3terminst}
\end{eqnarray}
where the angle brackets indicate where integrations  
have been carried out over time components of momenta.
This latter expression collapses to the 
impulse approximation form when  
Eq.~(\ref{eq:Salpeter}) is used in the first two terms:  
\begin{equation}
{\cal A}_{{\rm S},\mu} = \bar{\Gamma}'_{\rm S} \langle 
G_0' J_{\mu} G_0 \rangle  \Gamma_{\rm S}.
\label{eq:AmuET}
\end{equation}
The three terms in Eq.~(\ref{eq:3termcurrent})
are equal and they simplify to just one term 
in Eq.~(\ref{eq:AmuET}).   

However, in this work we did not begin with the ladder BSE. Instead we
began with the 4D equation (\ref{eq:4DET}). Constructing a conserved
impulse approximation current for the vertex function which is the
solution of Eq.~(\ref{eq:4DET}) is a little more involved. In
Ref.~\cite{PW98} we showed how to add a piece to the current
(\ref{eq:BSEIA}) which results in a conserved current when
$\Gamma$ is the solution of Eq.~(\ref{eq:4DET}), giving a total 4D current
\begin{equation}
  \bar{\Gamma}(P+q) G_0' J_{0,\mu} G_0 \Gamma(P) + 
                 \bar{\Gamma}(P+q) {G_C}_\mu \Gamma (P).
\label{eq:4DETAmu}
\end{equation}
The explicit expression for ${G_C}_\mu$ can be found in
Ref.~\cite{PW98}.

With this four-dimensional current in hand, we calculate 
the reduction to three dimensions as in Eq.~(\ref{eq:AmuET}).
Replacing $\Gamma$ by $\Gamma_{\rm ET}$ we end up with
\begin{equation}
{\cal A}_{{\rm inst},\mu}=\bar{\Gamma}'_{\rm ET}
\, {\cal G}_{{\rm inst},\mu}^\gamma  \, \Gamma_{\rm ET}
\label{eq:Amuinst}
\end{equation}
where:
\begin{equation}
{\cal G}_{{\rm inst},\mu}^\gamma = \langle G' J_{\mu} G \rangle 
= \langle G_0'J_{\mu}G_0 + G_{\rm C \mu}\rangle .
\end{equation}
This is analogous to the reduction employed for the bound-state
equation itself. Once again, this reduction can be performed in a
systematic fashion, but here we keep only the results for an
instantaneous interaction. In that case ${\cal A}_{{\rm inst},\mu}$ is
conserved, provided that $\Gamma_{\rm ET}$ is the solution of
Eq.~(\ref{eq:ET}) using the instant form of the interaction.  The
explicit form of ${\cal G}^{\gamma}_{\rm inst}$ is~\cite{PW98}:
\begin{eqnarray}
  {\cal G}_{{\rm inst},\mu}^\gamma({\bf p}_1,{\bf p}_2;P,Q)=i \,
  \langle d_1(p_1) \, d_2(p_2+Q) j^{(2)}_\mu d_2(p_2) \rangle
  \nonumber\\ + \, i \, \langle d_1(p_1) \, d_2^{\tilde{c}}(p_2+Q)
  j^{(2)}_{c,\mu} d_2^c(p_2) \rangle
 + (1 \leftrightarrow 2).
\label{eq:ETcurrent}
\end{eqnarray}
Only the $\gamma^\mu$ piece of
$j_\mu^{(n)}$ is relevant for charge conservation, since the piece
proportional to $\sigma^{\mu \nu}$ is automatically conserved.
Meanwhile, $d_n^c$ is a one-body Dirac propagator used in $G_C(P)$ to
construct the approximation to the crossed-ladder graphs.
Correspondingly, $d_n^{\tilde{c}}$ appears in $G_C(P+q)$. It does {\it
  not} equal $d_n^c$, even if particle $n$ is not the nucleon struck
by the photon.  Finally,

\begin{equation}
j^{(2)}_{c,\mu}= (q_2 \gamma^{(2)}_\mu - \tilde{j}^{(2)}_\mu);
\quad
  \tilde{j}^{(2)}_\mu=q_2 \frac{\hat{p}_{2 \mu}' + \hat{p}_{2
      \mu}}{\epsilon_2' + \epsilon_2} \gamma^{(2)}_0,
\label{eq:tildej2}
\end{equation}
with $\hat{p}_2=(\epsilon({\bf p}_2),{\bf p}_2)$.  The current defined
by Eqs.~(\ref{eq:Amuinst})--(\ref{eq:tildej2}) 
includes the effects of photons creating Z-graphs by means of 
couplings from positive-energy states to negative-energy states.  

This defines our impulse-approximation current.  Detailed results for
this current employed with the solutions of Eq.~(\ref{eq:ET}) were
presented in Ref.~\cite{PW98}.  The various refinements such as the
$j^{(2)}_{c,\mu} $ term in Eq.~(\ref{eq:ETcurrent}) that were not
considered by Hummel and Tjon~\cite{HT} produce only very small
changes for electron-deuteron scattering.  They are incorporated in
our calculations because they ensure current conservation when the ET
propagator is used.

\section{Dynamical boost}

\label{sec-boost}

To compute the current in Eq.~(\ref{eq:Amuinst}) we consider
electromagnetic matrix elements in the Breit frame where ${\bf P}=-{1
\over 2} {\bf q}$ in the initial state and ${\bf P}' = {1 \over 2}{\bf
q}$ in the final state.  In the impulse approximation, using the
instant interaction and the instant current in the Breit frame, the
current matrix element may be expressed as
\begin{eqnarray}
\int \frac{d^3 p}{(2\pi)^3} \bar{\Gamma}
({\bf p} + {1 \over 4}{\bf q}; {1 \over 2}{\bf q}) 
{\cal G}^\gamma_{{\rm inst}, \mu}
({\bf p}- {1 \over 2}{\bf q},-{\bf p};P,Q)
\nonumber \\
\Gamma ({\bf p} - {1 \over 4}{\bf q}; -{1 \over 2}{\bf q}).
\label{eq:BreitJmu}
\end{eqnarray}
In the ETIA calculation of Ref.~\cite{PW98} Eq.~(\ref{eq:BreitJmu}) is
calculated using vertex functions $\Gamma$ that are obtained by
solving the bound-state equation
\begin{equation} 
(E_{\bf P} - \epsilon_1 - \epsilon_2 )|E_{\bf P} \rangle = v |E_{\bf
P} \rangle,
\label{eq:bseq}
\end{equation}  
where $\epsilon_n = \sqrt{m^2 + {\bf p}_n^2}$, for the state vector
$|E_{\bf P} \rangle$, and then computing:
\begin{equation}
\Gamma({\bf k};{\bf P})=\langle {\bf k};{\bf P}|G_0^{-1}|E_{\bf P} \rangle.
\end{equation}
The operator 
\begin{equation}
H_{\rm ET}=\epsilon_1 + \epsilon_2 + v
\label{eq:HET}
\end{equation}
may be interpreted as the effective hamiltonian. With the usual
conditions on the interaction $v$, it is bounded from below and may be
used to generate a Hilbert space of states.

The interaction $v$ could be defined only in positive-energy states,
or it could be the effective interaction in positive-energy states
that includes the effects of couplings to negative-energy states: $+-,
-+$ and $--$. The couplings to negative-energy states obtained from
the ET reduction do not produce any singularities in the effective
interaction $v$.  Once the interaction is defined in the c.m. frame of
the two particles, where ${\bf P}=0$, the lowest eigenvalue is the
rest mass of the two-particle bound state, $M$.

The problem here is that this does not necessarily guarantee that
\begin{equation}
E_{\bf P} = \sqrt{M^2 + {\bf P}^2}
\label{eq:EP}
\end{equation}
in other frames.  That is because solving Eq.~(\ref{eq:bseq}) in an
arbitrary frame means using the ET interaction $K_{\rm ET}$ for $v$,
and $K_{\rm ET}$ is calculated using Dirac spinors with arguments
${\bf p} \pm {1\over 2}{\bf P}$. Such an interaction depends on total
momentum ${\bf P}$, and is {\it not} guaranteed to produce a bound
state of the appropriate energy.  In fact, in Ref.~\cite{PW98} we
found that $K_{\rm ET}$ must be renormalized by a factor $\lambda({\bf
P}^2)$ if (\ref{eq:EP}) is to hold in frames other than ${\bf
P}=0$. The {\it ad hoc} factor $\lambda$ varies linearly from 1 at
${\bf P}^2 = 0$ to about 1.10 at ${\bf P}^2 = 4$ GeV$^2$~\cite{PW98}.

This occurs because the two-body ET interaction obtained by solving
Eq.~(\ref{eq:bseq}) using the ET interaction does not respect the
Poincar\'e algebra.  The basic requirement of Poincar\'e invariance is
that states must transform under a unitary representation of the
Poincar\'e group.  The ten generators of translations in time,
translations in space, boosts, and rotations are the hamiltonian
operator, taken here to be of the form (\ref{eq:HET}), the total
momentum ${\bf P}$, the boost operator ${\bf K}$, and the angular
momentum operator ${\bf J}$.  They obey the following commutation
relations.
\begin{eqnarray}
\left[ {\bf P},  H \right] = 0,&\qquad& \left[ {\bf J},  H \right] = 0, 
\nonumber \\
\left[ {\bf K}, H \right] = i {\bf P}, &\qquad&
\left[  K_i, P_j \right] = i \delta_{ij} H ,   
\nonumber \\
\left[  J_i, P_j \right] = i \epsilon_{ijk} P_k ,&\qquad&
\left[  J_i, K_j \right] = i \epsilon_{ijk} K_k ,
\nonumber \\
\left[  J_i, J_j \right] = i \epsilon_{ijk} J_k ,&\qquad&
\left[  K_i, K_j \right] = -i \epsilon_{ijk} J_k .
\label{eq:algebra}
\end{eqnarray}

The two-body boost problem is to constrain the interaction, $v$, such
that the Poincar\'{e} generators satisfy the commutation rules that are
required for Poincar\'{e} invariance. Our strategy will be to take $v$
to agree with $K_{\rm ET}$ in the two-body rest-frame (${\bf P}=0$) and
then impose the conditions (\ref{eq:algebra}) in order to obtain
approximate expressions for the matrix elements of $v$ in a frame where
${\bf P} \neq 0$. We denote the hamiltonian of this ``approximate boost''
procedure by $H_{\rm ET}$.

As mentioned earlier, we could instead employ an exact solution
to the two-body boost problem by following Ref.~\cite{CP82} and
introducing the interaction $K_{\rm ET}({\bf P}=0)$ into the rest-frame
mass operator $\hat{M}$. A hamiltonian that satisfies the Poincar\'e
algebra would then be:
\begin{equation}  
H_{BT} = \sqrt{\hat{M}^2 + {\bf P}^2}.
\label{eq:BT}
\end{equation} 
 
The reason we choose not to apply Eq.~(\ref{eq:BT}) is that for
electromagnetic interactions we also need current operators that are
consistent with both the rest-frame and boosted hamiltonian.  The ET
formalism solves that problem by performing a consistent reduction of
the quantum field theory. As discussed above, this yields currents
which are consistent with $K_{\rm ET}$. Such a construction is
significantly more difficult for $H_{BT}$. To indicate this we merely
point out that if $H_{\rm ET}$ exactly satisfied the Poincar\'e
algebra then the two different hamiltonians $H_{\rm ET}$ and
$H_{BT}$ would be related by a unitary transformation, i.e., there
would exist an operator $U$ with $U^\dag U=1$ and:
\begin{equation}
H_{\rm ET}= U H_{BT} U^{\dag}.
\label{eq:unitary}
\end{equation} 
If this $U$ were known, we could derive currents for use with states
obtained from the hamiltonian (\ref{eq:BT}) by transforming the
currents derived in the ET formalism using the transformation
(\ref{eq:unitary}). However, this unitary transformation is {\it not}
known. So, in order to respect the consistency between the ET bound
states and current operators, we will focus on the problem of finding
an approximate boost that preserves the form of $H_{\rm ET}$ in frames
with ${\bf P} \neq 0$.  

It should be mentioned that if ${\bf P} \neq 0$ the 
hamiltonian $H_{\rm{ET}}$  
has small differences from the operator 
$\epsilon_1 + \epsilon_2 + K_{\rm{ ET}}$ 
that was used in our analysis of the Ward-Takahashi identity. 
We leave the issue of boosting the currents so as to exactly
maintain the Ward-Takahashi identity at ${\bf P} \neq 0$ as 
an unsolved problem.  

An approximate solution to the boost problem within the subspace of an
eigenvalue of the hamiltonian has been obtained in Ref.~\cite{Wa01}
for the case of two scalar particles.  It provides an exact result for
the energy of the two-body system.  Here we review the basic elements
of that proof and extend the result approximately to the case of two
spin-half particles.

For the ET (and a number of other 3D) formalism(s), the two-body boost
problem is the same as for the instant form of hamiltonian
dynamics~\cite{Dirac49}.  In this form of dynamics, total momentum and
angular momentum operators do not depend on the interactions.  They
are:
\begin{eqnarray}
{\bf P} &=& {\bf p}_1 + {\bf p}_2 ,\\
{\bf J} = {\bf r}_1 \times {\bf p}_1 &+& {\bf r}_2 \times {\bf p}_2  + 
{ 1\over2}(\sigma_1 +\sigma_2). 
\end{eqnarray}
It follows from $H$ commuting with ${\bf P}$ and ${\bf J}$ that $v$
must be translationally and rotationally invariant.  The commutation
relation of ${\bf K}$ and $H$ requires the boost operator to depend on
the interaction.  Such a boost is called ``dynamical'' in order to
distinguish it from a ``kinematical'' boost for which the boost generator
does not involve the interaction.  The commutation rule between the
dynamical boost generator and hamiltonian thus involves $v^2$ terms.

Bakamjian and Thomas~\cite{BT53} derived the form of the boost
operator for instant dynamics.  For a free boost, the operator is
\begin{equation}
{\bf K}_0 = {1 \over 2} \left( {\bf r}_1 \epsilon_1 + \epsilon_1 {\bf
  r}_1 \right) -{1 \over 2} \frac{\sigma_1 \times {\bf p}_1}{\epsilon_1
  + m} + (1 \rightarrow 2 ) .
\end{equation}
When interactions are present, there is an interaction part of the
hamiltonian, $v$, and an interaction part of the boost
operator that is given approximately by
\begin{equation}
{\bf K}_v = {1 \over 2} \left( {\bf R} v  + v {\bf R} \right),
\label{eq:Kboost}
\end{equation}
where ${\bf R} = {1 \over 2}({\bf r}_1 + {\bf r}_2)$.  This boost
operator omits some terms from the standard Newton-Wigner
construction~\cite{CP82} for ${\bf R}$, and likely could be improved.
We leave such possible improvement of the boost for future analysis.
Although it involves a significant approximation, in the case of the
deuteron the boost operator of Eq.~(\ref{eq:Kboost}) yields the
relationship (\ref{eq:EP}) between the mass and energy of the bound
state to good accuracy. 

With the definitions given above, $[K_i,P_j] = i H \delta_{ij}$ is
satisfied.  However, there is an error term of order $1/m^2$
in the commutation relation $[K_i,K_j] = -i\epsilon_{ijk}J_k$.  
Accepting this error, the interaction $v$ must take an
appropriate, but unknown, form consistent with $[{\bf K},H] = i{\bf P}$.

Noting that the free-boost operator, ${\bf K}_0$, and the free
hamiltonian, $H_0 = \epsilon_1 + \epsilon_2$, obey the commutation
relation $[K_0, H_0] = i{\bf P}$, it follows that when
interaction-dependent terms are introduced into that commutation
relation, their contributions must sum to zero, i.e.,
\begin{equation}
[{\bf K}_0 , v] + [{\bf K}_v, H_0] + [{\bf K}_v, v] = 0.
\end{equation}
This is equivalent to 
\begin{eqnarray}
\frac{1}{2}[{\bf R}, H_0 v + v H_0 + v^2] 
+ \frac{1}{4} [ {\bf r} \Delta + 
\Delta {\bf r}, v] 
+ [ {\bf K}_{\sigma} , v]= 0,\nonumber
\end{eqnarray}
where we have defined:
\begin{eqnarray}
 {\bf K}_{\sigma} &\equiv& -{1 \over 2} \frac{\sigma_1 \times {\bf p}_1}{\epsilon_1 + m} + (1  \rightarrow 2 );\\
\Delta &\equiv& \epsilon_1 - \epsilon_2.
\end{eqnarray}
Algebraic manipulations lead to
\begin{eqnarray}
[{\bf R}, H_0] v + v [{\bf R}, H_0] +{1 \over 2} H [{\bf R}, v]
+ {1 \over 2} [{\bf R}, v] H \nonumber \\
+ {1 \over 2} (\Delta {\bf r} v - v {\bf r} \Delta)
 +[ {\bf K}_{\sigma} , v]= 0.
\label{eq:comm}
\end{eqnarray}

In quantum field theory the boost of a mass eigenstate appears to be
kinematical. In fact the boost velocity $\beta = {\bf P}/E_{{\bf P}}$
depends upon the eigenvalue in the c.m. frame, $M$, which of course
involves the interaction.  This simple observation provides the key to
solving the boost problem in instant dynamics.  The $v^2$ terms in the
dynamical boost may be eliminated when the boost is considered within
the subspace of a single eigenvalue of the mass.

The boost velocity appropriate to a mass eigenstate does not enter the
commutation relations.  In order to restrict the boost to the subspace
corresponding to a particular eigenvalue of $H$ we evaluate the
matrix element of (\ref{eq:comm}) between eigenstates of mass $M$
that obey Eq.~(\ref{eq:bseq}).  This yields:
\begin{eqnarray} 
\langle E_{{\bf P}}| \Big( [{\bf R}, H_0] v + v [{\bf R}, H_0] 
+ {1 \over 2} E_{{\bf P}} [{\bf R}, v] + 
{1 \over 2}[{\bf R}, v]E_{{\bf P}}  \nonumber \\
+ {1 \over 2} (\Delta {\bf r} v - v {\bf r} \Delta)
+ [ {\bf K}_{\sigma} , v] \Big)
|E_{{\bf P}}\rangle = 0 .
\label{eq:CR1}
\end{eqnarray}
The $v^2$ pieces in the third and fourth terms of Eq.~(\ref{eq:comm})
have been eliminated in favor of the energy eigenvalue
$E_{{\bf P}}$.  Because of this, the boost appears to be kinematical
in much the same way that the boost of a mass eigenstate in 
quantum field theory appears to be kinematical.  

Performing the manipulations discussed in Ref.~\cite{Wa01} we can
convert Eq.~(\ref{eq:CR1}) into an equation that is linear in $v$,
\begin{eqnarray}
\langle E_{{\bf P}}|\Big\{  \frac{1}{2} \left( 1 + \frac{H_0}{ E_{{\bf P}}} \right) 
[{\bf R}, H_0] v +  \frac{1}{2} v [{\bf R}, H_0] 
\Big( 1 + \frac{H_0}{ E_{{\bf P}}} \Big) \nonumber \\
+  {1 \over 2} E_{{\bf P}} [{\bf R}, v]  + 
{1 \over 2}[{\bf R}, v]E_{{\bf P}} +
\frac{ {\bf p} \cdot{\bf P} }{E_{{\bf P}}}{\bf r} v
- v {\bf r}\frac{ {\bf p} \cdot{\bf P} }{E_{{\bf P}}} \nonumber \\
+ [ {\bf K}_{\sigma} , v] 
\Big\} |E_{{\bf P}}\rangle = 0  .
\label{eq:CR3}
\end{eqnarray} 

This equation is solved in momentum space in order to determine the
form of $v$.  Momentum-space matrix elements involve
\begin{eqnarray}
&&\int \frac{d^3p'}{(2\pi)^3}\frac{d^3p}{(2\pi)^3} 
\langle E_{{\bf P}} | {\bf p}'; {\bf P}\rangle E_{{\bf P}}
\nonumber \\
&&\qquad \Big\{ 
\left[ {\bf A}({\bf p}'; {\bf P})  +
{\bf A}({\bf p}; {\bf P}) + {\bf B_{\rm op}}\right] 
\langle{\bf p}';{\bf P} | v | {\bf p};{\bf P}\rangle   
\nonumber \\
&& \qquad \qquad -i [ {\bf K}_{{\bf P},\sigma} , \langle{\bf p}';{\bf P} | v | 
{\bf p};{\bf P}\rangle ] -i {\bf k}'_{\sigma}
\langle{\bf p}';{\bf P} | v | {\bf p};{\bf P}\rangle 
\nonumber \\
&&\qquad \qquad \qquad + i\langle{\bf p}';{\bf P} | v |{\bf p};{\bf P}\rangle {\bf k}_{\sigma}
\Big\} \langle {\bf p};{\bf P} | E_{{\bf P}}\rangle   
= 0, 
\label{eq:pspace}
\end{eqnarray}
where  
\begin{equation}
{\bf A}({\bf p};{\bf P}) = \frac{1}{2E_{{\bf P}}} \left( 1 +
\frac{d({\bf p};{\bf P})}{E_{{\bf P}}} \right) \frac{\partial d({\bf
p};{\bf P})}{\partial {\bf P}},
\end{equation}
with $d \equiv \epsilon_1({\bf p};{\bf P}) + \epsilon_2({\bf p};{\bf
P})$ being the value of $H_0$, and $A({\bf p}'; {\bf P})$ the same
as $A({\bf p}; {\bf P})$, except with ${\bf p} \rightarrow {\bf p}'$.
Meanwhile,
\begin{equation}
{\bf B}_{\rm op} \equiv 
\frac{\partial }{\partial {\bf P}} 
+ \frac{{\bf p}'\cdot{\bf P}}{E_{{\bf P}}^2} 
\frac{\partial}{\partial{\bf p}'}  
+ \frac{{\bf p}\cdot{\bf P}}{E_{{\bf P}}^2} 
\frac{\partial}{\partial{\bf p}},
\label{eq:Bop}
\end{equation}
\begin{equation} 
{\bf K}_{{\bf P},\sigma} = - \frac{1}{2} \frac{(\sigma_1+\sigma_2) \times {\bf P}}
{E_{\bf P} (E_{\bf P} + M)} ,
\end{equation}
and 
\begin{equation}
{\bf k}_{\sigma} = - \frac{1}{2} \frac{\sigma_1 \times {\bf q}_1}
{E_{\bf P}(\epsilon_1 + m)} 
+ ( 1 \rightarrow 2) ,
\end{equation} 
with 
\begin{equation}
{\bf q}_1 = {\bf p} - \left( \frac{1}{2} - \frac{\epsilon_1 + m}
{E_{\bf P} + M}\right) {\bf P}.
\end{equation}
Derivatives in ${\bf B}_{\rm op}$ act only on the interaction.  

With spin effects omitted, a solution of Eq.~(\ref{eq:pspace}) was
obtained such that
\begin{equation}
\langle{\bf p}';{\bf P} | v |{\bf p};{\bf P}\rangle =
f({\bf p};{\bf P}) \tilde{v}({\bf p}',{\bf p};{\bf P})
f({\bf p'};{\bf P}).
\label{eq:piint}
\end{equation}
Here:
\begin{equation}
(A({\bf p}; {\bf P})+ {\bf B}_{\rm op})f({\bf p};{\bf P}) = 0.
\label{eq:feq}
\end{equation}

The form of $\tilde{v}$ is deduced from the condition ${\bf B}_{op}
\tilde{v} = 0$ and the boundary condition that, for ${\bf P} = 0$, the
interaction must be the c.m.~frame one.
Now constructing the rotational scalars   
\begin{eqnarray}
{\bf p}_c^2 &=& {\bf p}^2 - \frac{({\bf p}\cdot {\bf P})^2}{E_{{\bf P}}^2} ,
\nonumber \\
{\bf p}_c^{'2} &=& {\bf p}^{'2} - \frac{({\bf p}'\cdot {\bf P})^2}{E_{{\bf P}}^2} ,
\nonumber \\
{\bf p}_c\cdot {\bf p}_c' &=& {\bf p}\cdot {\bf p} - 
\frac{({\bf p}\cdot {\bf P})({\bf p}'\cdot {\bf P})}{E_{{\bf P}}^2}, 
\label{eq:pc2}
\end{eqnarray}
one can check that 
\begin{eqnarray}
{\bf B}_{op} {\bf p}_c^2 = 0, \nonumber \\
{\bf B}_{op} {\bf p}_c^{'2} = 0, \nonumber \\
{\bf B}_{op} {\bf p}_c\cdot {\bf p}'_c = 0.
\end{eqnarray}
So, it follows that if $\tilde{v} = v_c({\bf p}_c', {\bf p}_c)$ is an
arbitrary function of ${\bf p}_c^2$, ${\bf p}_c^{'2}$ and ${\bf p}_c 
\cdot {\bf p}'_c$, then
\begin{equation}
{\bf B}_{op} v_c({\bf p}_c', {\bf p}_c) =  0,
\end{equation}
and the condition ${\bf B}_{op} \tilde{v}=0$ is satisfied.
Thus, in the absence of spin,
\begin{equation}
\tilde{v}({\bf p}',{\bf p};{\bf P}) = v_c({\bf p}_c',{\bf p}_c).
\label{eq:vtildevc}
\end{equation}
where 
\begin{equation}
{\bf p}_c \equiv {\bf p} - \frac{({\bf p}\cdot {\bf P}) {\bf P}}
{E_{{\bf P}}( E_{{\bf P}} + M) },
\label{eq:pc}
\end{equation}
and  
\begin{equation}
{\bf p}'_c \equiv {\bf p}' - \frac{({\bf p}'\cdot {\bf P}) {\bf P}}
{E_{{\bf P}}( E_{{\bf P}} + M) } .
\label{eq:pc'}
\end{equation}
In the c.m. frame, ${\bf p}_c$ and ${\bf p}'_c$ are the standard
relative momenta.  When the total momentum is in the $z$ direction,
the $z$-component of relative momentum ${\bf p}_c$ is contracted
according to $p_{cz} = p_z/\gamma$, where $\gamma = E_{{\bf P}}/M$.
The components of relative momenta perpendicular to the total momentum
are unaffected: ${\bf p}_{c\perp} = {\bf p}_{\perp}$.  The same rule
applies to ${\bf p}_c'$.

Solving Eq.~(\ref{eq:feq}) for $f({\bf p};{\bf P})$ subject to the boundary
condition that $f({\bf p},0)=1$ we find:
\begin{eqnarray}
f^2({\bf p};{\bf P}) = \frac{M}{E_{{\bf P}}} \left(\frac{ E_{{\bf P}} 
- \epsilon_1({\bf p}; {\bf P}) - \epsilon_2({\bf p}; {\bf P})  }
{M - 2 \epsilon({\bf p}_c;0)} \right).
\label{eq:fsol}
\end{eqnarray}
This completes the proof that the 
commutator relation $[{\bf K},H] = i{\bf P}$ is satisfied exactly
in the subspace of eigenvalue $E_{{\bf P}}$ when spin 
effects are omitted.

When spin effects are included, we find that if the c.m. frame 
interaction
is expressed in terms of a matrix element involving Dirac
spinors depending upon ${\bf p}_c$ and ${\bf p}_c'$,
then the commutator term involving ${\bf K}_{{\bf P},\sigma}$ 
and the spin terms generated by ${\bf B}_{op}$ acting on
the Dirac spinors in Eq.~(\ref{eq:pspace}) cancel. That is, let
\begin{eqnarray}
v_c({\bf p}_c',{\bf p}_c) = \bar{u}_1({\bf p}_c')
\bar{u}_2({\bf p}_c')\  \hat{v}({\bf p}_c',{\bf p}_c) 
\ u_1({\bf p}_c) u_2({\bf p}_c)
\label{eq:vcspin}
\end{eqnarray}
where $\hat{v}({\bf p}_c',{\bf p}_c)$ is a rotationally invariant
function of its arguments that may involve local Dirac matrix
structures, for example the Fermi covariants. The Dirac spinors $u_1$,
$u_2$, etc. in (\ref{eq:vcspin}) omit the usual Pauli spinor factors,
i.e., they are boost operators for Dirac spinors similar in form to
Eq.~(\ref{eq:ustar}) below.  It follows that the spinor matrix element
in (\ref{eq:vcspin}) involves products of terms like $\sigma_1 \cdot
{\bf p}_c$ and $\sigma_2 \cdot {\bf p}_c$ from the spinors and
$\hat{v}$. Now consider the Dirac matrix structure $\Gamma_1 \Gamma_2$
that arises from these products. The cancellation we want will
occur, provided that
\begin{equation}
{\bf B}_{\rm op} \Gamma_1 \Gamma_2 -i [{\bf K}_{{\bf P}\sigma},
\Gamma_1\Gamma_2] = 0.
\end{equation}
But the Dirac structures $\Gamma_1 \Gamma_2$ either involve $\sigma_1
\cdot \sigma_2$ or---as in the case of $\pi N N$ coupling---factors of
$\sigma_n \cdot {\bf p}_c$.  A straightforward calculation then shows
that,
\begin{equation}
{\bf B}_{op} \sigma_1 \cdot {\bf p}_c -i [ {\bf K}_{{\bf P},\sigma}, 
\sigma_1 \cdot {\bf p}_c ] = 0,
\end{equation}
and that a similar identity holds also for argument ${\bf p}_c'$ and
spin operator $\sigma_2$. Since ${\bf K}_{{\bf P},\sigma}$ commutes
with $\sigma_1 \cdot \sigma_2$, it follows that:
\begin{equation}
{\bf B}_{op} v_c({\bf p}_c',{\bf p}_c) -i[ {\bf K}_{{\bf P},\sigma}, v_c({\bf p}_c',{\bf p}_c)] = 0.
\end{equation} 

The remainder of the boost operator can be understood as requiring
a rotation of spins when ${\bf P} \neq 0$,  
\begin{equation}
R_i({\bf p}) \approx 1 + \frac{i {\bf P}\cdot \sigma_i 
\times {\bf p}}{2(\epsilon_c + m)^2} + \cdots ,
\label{eq:Rspin}
\end{equation}
where $i=1$ or 2 and omitted terms are of order $1/m^4$.~\cite{m4}  That is, 
the interaction in a frame with nonzero total momentum 
should have the final form 
\begin{eqnarray}
\langle{\bf p}';{\bf P} | v |{\bf p};{\bf P}\rangle 
 = f({\bf p}';{\bf P})R_1' R_2' \bar{u}_1({\bf p}_c')
\bar{u}_2({\bf p}_c') \nonumber \\
\hat{v}({\bf p}_c',{\bf p}_c) 
u_1({\bf p}_c) u_2({\bf p}_c) R_1 R_2 f({\bf p};{\bf P}).
\label{eq:vPspin}
\end{eqnarray}
Factors $R_i'$ and $R_i$ allow $1/m^2$ parts
of the terms involving ${\bf k}'_{\sigma}$ and ${\bf k}_{\sigma}$
to be canceled by new terms generated when ${\bf B}_{op}$ acts on the 
spin rotation factors in Eq.~(\ref{eq:pspace}).  
However, there are order $1/m^4$ errors that cannot
be canceled: we accept these as error terms of the boost related
to the fact that our boost generator is not exact.

In summary, we find that if an arbitrary c.m. frame interaction that
is rotationally and translationally invariant is constructed in other
frames according to Eq.~(\ref{eq:vPspin}), then a bound state of mass $M$
defined by solution of the c.m. frame equation,
\begin{equation}
[M - 2 \epsilon({\bf p}';0) ]\Psi_c({\bf p}') = \int \frac{d^3p}{(2\pi)^3}
v_c({\bf p}',{\bf p}) \Psi_c({\bf p}),
\end{equation}
corresponds in another frame to a state of energy $E_{\bf P}$, given
by solving the eigenvalue equation:
\begin{eqnarray}
&&[E_{{\bf P}} - \epsilon_1({\bf p}';{\bf P}) 
- \epsilon_2({\bf p}';{\bf P})] \Psi({\bf p}';{\bf P}) = 
\nonumber \\ 
&& \qquad \qquad \int \frac{d^3p}{(2\pi)^3} 
\langle {\bf p}',{\bf P}|v|{\bf p};{\bf P}\rangle \Psi({\bf p};{\bf P}).
\label{eq:BreitEq}
\end{eqnarray}

Wave functions for the initial and final state are now obtained from
Eq.~(\ref{eq:BreitEq}) using the interaction defined as in
Eq.~(\ref{eq:vPspin}).  This provides wave functions of energy $E_{\bf
P}$ that correspond to a mass $M$ in the c.m. frame of the
two-nucleons, and which can be inserted into Eq.~(\ref{eq:BreitJmu}).
The spin rotation factors are given approximately by
Eq.~(\ref{eq:Rspin}), within errors of order $1/m^4$. 
We have omitted spin rotation effects in our calculations but we find
the correct relation between energy and mass with high accuracy.
The boost rule just stated is approximate because we have
omitted terms that would be required in order to achieve an exact
solution of the Poincar\'{e} algebra.  It is designed to be consistent with
the currents derived using the ET reduction.

\section{Pseudovector-pion contribution to the isoscalar two-body four
current}

The derivation presented in Sec.~\ref{sec:current-conservation} 
involves a straightforward reduction of conserved
four-dimensional matrix elements of the electromagnetic current.  It
neglects retardation effects (known to be small) but also neglects the
dependence on the relative energy $p_0$ of meson-nucleon vertex
functions.

\begin{figure}[h,t,b]
\parbox{10cm}{\resizebox{10cm}{!}{\includegraphics*[2cm,7cm][21.5cm,23cm]
{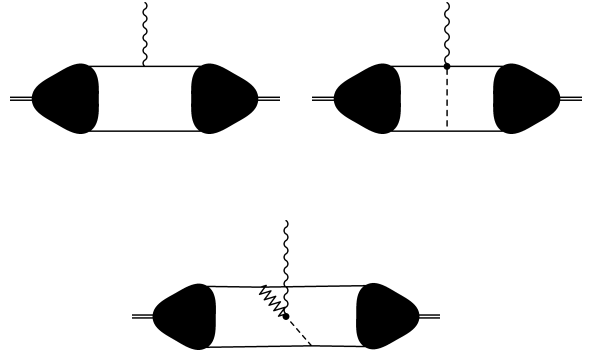}}}
\vskip -3.0cm
\caption{On the left of the upper line we depict the
  positive-energy-state impulse-approximation mechanism for
  electron-deuteron scattering. On the right of that line is the
  PV-contact MEC. In the bottom line we show the $\rho \pi \gamma$
  MEC, with the $\rho$-meson represented by the wavy line. In each
  case the blobs are the deuteron vertex functions $\Gamma$.}
\label{fig-current}  
\end{figure}

As we will show here, keeping the $p_0$-dependence of the $\pi NN$
vertex results in an additional contribution to the 3D current that
can be interpreted as a ``meson-exchange current'' contribution to the
current operator.  As stressed by Friar~\cite{Friar} and Adam and
Arenh\"ovel~\cite{AA97}, it is critical that these MEC contributions be
evaluated in a manner consistent with that used to evaluate the $NN$
interaction. Here we implement this consistency by recalling that the
instant $NN$ potential $K_{\rm ET}$ is derived by computing the ET
matrix elements using a four-dimensional kernel and then taking the
static limit $|E - \epsilon - \epsilon'| \ll \omega$ in the
denominator.  Thus in order to obtain a consistent ${\cal A}_\mu$ for
pseudovector pion coupling we should implement
Eq.~(\ref{eq:3terminst}) with the 4D $K$ given by Eq.~(\ref{eq:4DKpi})
and also take the static limit there.  This leads to a pion-range
contribution to the isoscalar four current, one first identified by
Riska a number of years ago. This contribution was {\it not} included
in calculations of Ref.~\cite{PW98} and its omission is part of the
reason that the data for $A$ is underpredicted there.
It is depicted on the right-hand side of the upper line 
of Fig.~\ref{fig-current}.
For pseudovector $\pi N$ coupling, the extra current is derived from
Eq.~(\ref{eq:3termcurrent}) by evaluating the expressions using the
positive-energy part of each Dirac propagator and the $\pi$-exchange
kernel of Eq.~(\ref{eq:4DKpi}).  Following the procedure
outlined above Eq.~(\ref{eq:KpiET}) leads to an expression very
similar to Eq.~(\ref{eq:3terminst}), as follows
\begin{eqnarray}
&& {\cal A}_\mu=\bar{\Gamma}'_{\rm ET} \{ 
\langle G_0' J_{\mu} G_0 \rangle  K_{\rm ET}^{\rm A,{\pi}} \langle  G_0 \rangle +
\langle G_0' \rangle K_{\rm ET}^{\rm B,{\pi}} \langle  G_0' J_{\mu} G_0 \rangle  
\nonumber \\ 
&&\qquad \qquad \qquad - \langle G_0' J_{\mu} G_0 \rangle 
\} \Gamma_{\rm ET}.
\label{eq:3terminstAB}
\end{eqnarray}
where 
\begin{eqnarray}
&&K^{\rm A,{\pi}}_{\rm ET} =  K^{\rm B,{\pi}}_{\rm ET}=\nonumber\\
&&\qquad 
- \frac{g_{\pi}^2\tau_1\cdot\tau_2}{{\bf q}^2 + m_{\pi}^2} \bar{u}_1'\bar{u}_2'
\Big\{ \gamma_1^5 + \frac{-r_A -\epsilon_1 + \epsilon'_1}{2m} 
\gamma_1^5 \gamma_1^0 \Big\} \nonumber \\
&&\qquad \qquad
\Big\{ \gamma_2^5 + \frac{r_A -\epsilon_2 + \epsilon'_2}{2m} 
\gamma_2^5 \gamma_2^0 \Big\}u_1 u_2,
\label{eq:KApi}
\end{eqnarray}
with $r_A = {1 \over 2} ( E_{\bf P} - 2 \epsilon_2 + \epsilon_2' -
\epsilon_1')$.  The factors in curly brackets in Eq.~(\ref{eq:KApi})
are similar to those in Eq.~(\ref{eq:KpiET}). The
replacement of $p_1^0 - {p_1'}^0$ by $-r_A$ occurs as the result of
performing the integrations over time-components of momenta as
dictated by the reduction to three dimensions. Conversely
$p_2^0 - {p_2'}^0$ gets replaced by $+r_A$.  

The extra terms that arise due to the retention of $p'_0 -p_0$ pieces
in the numerator of the pion exchange can now be determined by
comparing Eq.~(\ref{eq:3terminstAB}) with Eq.~(\ref{eq:3terminst}).
This leads to the following expression, which is what we calculate
below in order to include the PV-contact MEC:
\begin{eqnarray}
&&{\cal A}_{\rm MEC,\mu}=\bar{\Gamma}'_{\rm ET} \Big\{ \langle G_0' J_{\mu} G_0
\rangle \nonumber\\
&&\qquad + \langle G_0'J_{\mu} G_0 \rangle \left( K_{\rm ET}^{\rm A,{\pi}}
- K_{\rm ET}^{\pi} \right) \langle G_0 \rangle \nonumber \\ 
&& \qquad + \langle G_0'
\rangle \left( K_{\rm ET}^{\rm B,{\pi}} - K_{\rm ET}^{\pi} \right)
\langle G_0'J_{\mu} G_0 \rangle \Big\} \Gamma_{\rm ET}.
\label{eq:PVMEC}
\end{eqnarray}

In fact, the dominant contribution here is due to $J_0$. Contributions
to $\vec{J}$ are suppressed by $p/M$ and are numerically small, but
they are included automatically in our formalism.
The resultant expression for the two-body piece of $J_0$
agrees with Adam {\it et al.}~\cite{Ad93} at leading order in
$p/m$. We have also checked that the expression (\ref{eq:PVMEC}) for
this PV-coupling current agrees numerically with the more conventional
expression as a two-body operator---again, to leading order in $p/m$.

\section{The $\rho \pi \gamma$ exchange current}

Because of the quantum numbers of the deuteron, the $\rho \pi \gamma$
meson-exchange current is generally thought to be the lowest-mass
mesonic excitation that makes a contribution to the electromagnetic
deuteron current.  The $\rho \pi \gamma$ exchange current can be used
to provide a resonance-saturation model of the two-body contribution
to $\vec{J}$. This meson-exchange piece of $G_M$ is of $O(e{\cal
P}^4)$, and has an unknown coupling from ${\cal L}_{\pi
N}^{(3)}$. The value of this unknown coupling can be modeled
by assuming it is dominated by the $\rho$ excitation, see
e.g. Ref.~\cite{Pa95}.

The Lagrangian governing the $\rho \pi \gamma$ vertex is:
\begin{equation}
  {\cal L}\, =-e \, \frac{g_{\rho \pi \gamma}}{2 m_\rho} \,
  \epsilon_{\alpha \beta \gamma \delta} F^{\alpha \beta} \,
  \vec{\rho}^{\, \gamma} \cdot \partial^\delta \vec{\pi}, 
\end{equation}
while the $\rho NN$ vertex has the form
\begin{equation}
{\cal L}_{\rho NN}=\bar{N}  g_{\rho}\Big( \gamma^{\mu} + i\frac{f_{\rho}}
{g_{\rho}}  \frac{\sigma^{\mu \nu}q_{\nu}}{2m} \Big)N.
\end{equation}

This yields the two-body current depicted on the lower line of
Fig.~\ref{fig-current}. Note that this current is automatically
conserved. In calculating it we use the same vertex functions and form
factors employed at the $\pi N$ and $\rho N$ vertices in calculating
the $NN$ potential.  The coupling $g_{\rho \pi \gamma}=0.563$ is
extracted from the decay $\rho \rightarrow \pi \gamma$.  This
value was used by Hummel and Tjon~\cite{HT}, however, 
Truhl\'{i}k, Smejkal and Khanna~\cite{TSK} recently have calculated somewhat higher values,
namely .585 ( based on $\rho^{\pm}$ decay) and .610 
( based on $\rho^0$ decay).
Although the overall sign of $g_{\rho\pi\gamma}$ is not 
determined by the experimental data, a recent lattice 
QCD calculation gives
a positive value~\cite{Ed04}. Quark models 
such as the one of Ref.~\cite{IG93} also provide 
a positive value for $g_{\rho\pi\gamma}$.
We have used $g_{\rho\pi\gamma}=$.563 in our calculations.  
For consistency with the Bonn-B potential the ratio $f_{\rho}/g_{\rho}$
was initially chosen to be 6.1.  However, $f_\rho/g_\rho$ is not well
determined by fitting the $NN$ data and recent work has suggested that
a smaller value $f_{\rho}/g_{\rho} \approx 4.5$ is 
acceptable~\cite{Co04}. This would be closer to the 
$f_\rho/g_\rho \approx 3.7 $ value obtained when vector-meson dominance 
is used to explain the isovector anomalous magnetic moment
of the nucleon~\cite{Bhad}.
We note that in nonrelativistic calculations
the $f_{\rho}/g_{\rho}$ term often is neglected altogether on the
grounds that it is higher order in $1/m$~\cite{Wi95}. In contrast,
Hummel and Tjon performed relativistic calculations of deuteron form
factors using $f_{\rho}/g_{\rho}=6.8$. They found that the
contribution of the tensor $\rho N$ coupling to $G_M$ cancels with the
contribution of the vector $\rho N$ coupling when $|{\bf q}|$ is about
1 GeV. This throws into question the validity of neglecting the tensor
piece of the $\rho NN$ coupling if $f_\rho/g_\rho$ is as large as this.

Even were $f_\rho/g_\rho$ known precisely there is still another
significant source of uncertainty in evaluating this exchange current:
its contribution to electron-deuteron scattering depends crucially on
the behavior of the current operator as a function of $Q^2$, and hence
on the $\rho \pi \gamma$ form factor. In the work of Hummel and Tjon
vector-meson dominance was used to obtain a $\rho \pi \gamma$ form
factor given solely by the $\omega$ meson:
\begin{equation}
F_{\rho \pi \gamma}(Q^2)=\frac{1}{Q^2 + m_\omega^2}.
\label{eq:rpgf}
\end{equation}
This same $\rho \pi \gamma$ form factor is also employed in the
non-relativistic calculations of Refs.~\cite{Wi95,Schiavilla02}. Other
calculations have used form factors based on quark
models~\cite{vO95,IG93}. Such form factors tend to reduce the contribution
of this MEC, which is also very sensitive to the cutoff masses in the
$\pi N$ and $\rho N$ vertices.

No experimental data are available to constrain the form factor that
should be used for the $\rho \pi \gamma$ vertex.  Lattice QCD
calculations can determine the form factor for on-shell $\pi$ and
$\rho$ mesons.  Recent lattice calculations of Edwards~\cite{Ed04}
indicate that for $Q^2$ up to 0.6 GeV$^2$ the $\rho\pi\gamma$
for factor agrees with the vector-meson dominance form (\ref{eq:rpgf}).

\section{Z-graphs}

In the full ET analysis the interactions and propagators are defined
on a complete set of Dirac plane-wave states: $++$, $+-$, $-+$ and
$--$.  The bound-state equation is solved with all possible couplings
between these states, and the current matrix elements allow for all
possible transitions between the states.  Physically, coupling to
negative-energy states allows positive-energy, plane-wave solutions of
the free Dirac equation to be distorted in the presence of a meson
field.  The simplest example of this effect---and one of the
motivations for including it---is the shift of the mass of a Dirac
particle in a uniform external scalar field: $ m \rightarrow m^* =
m+S$.  Positive-energy solutions of the Dirac equation in this
scalar field have the form
\begin{equation}
u(p) \approx \pmatrix{ 1 \cr \frac{\sigma \cdot {\bf p}}{2m^*} }.
\label{eq:ustar}
\end{equation}
To recapture the effect of the replacement $m \rightarrow m^*$
order-by-order in $S$ the coupling of positive-energy states of mass
$m$ to negative-energy states must be included in the calculation.  So
allowing for couplings between all positive- and negative-energy
components---as we do in the full ET formalism---is a general way to
incorporate this kind of distortion of the Dirac spinors in the
theory.

It is interesting to explore the extent to which these effects of
coupling to negative-energy states are captured if we include them
only via first-order perturbation theory in the nucleon-nucleon
potential. Thus, instead of solving (\ref{eq:ET}) in all $\rho$-spin
sectors, $\rho_1, \rho_2=++,+-,-+,--$, we solve it in $++$ states
alone to obtain $\Gamma_{\rm ET}^{++}$ and then generate the couplings
to negative-energy states by treating the difference between the full
interaction and the $++$ states-only interaction in first-order 
perturbation theory. This
leads to the following expression for the current matrix element:
\begin{eqnarray}
&&{\cal A}_{\rm Z,\mu}=\bar{\Gamma}^{' ++}_{\rm ET} \Big\{ \langle G_0' J_{\mu} G_0
\rangle \nonumber\\
&&\qquad \qquad + \langle G_0'J_{\mu} G_0 \rangle \left( K_{\rm ET}
- K_{\rm ET}^{++,++} \right) \langle G_0 \rangle \nonumber \\ 
&& \qquad  + \langle G_0'
\rangle \left( K_{\rm ET} - K_{\rm ET}^{++,++} \right)
\langle G_0'J_{\mu} G_0 \rangle \Big\} \Gamma_{\rm ET}^{++}.
\label{eq:Zgraphs}
\end{eqnarray}
The second and third terms in Eq.~(\ref{eq:Zgraphs}) provide the
leading-order corrections due to negative-energy components in the
initial and final states, respectively.  The leading-order Z-graph
calculations we present below are performed in this way.

\section{Results}

\label{sec-results}

With all the theoretical pieces of the puzzle assembled we now
calculate electron-deuteron scattering observables. The vertex functions employed
are the ones calculated with all positive and negative-energy states
included, as described in Ref.~\cite{PW98}.  If the negative-energy
states are dropped the interaction is exactly the Bonn-B potential for
the Thompson equation, as derived and fitted to $NN$ phase shifts in
Ref.~\cite{Ma89}. This model gives a reasonable fit to the $NN$ data,
and good deuteron static properties, although it is not as good a fit
as some more recent $NN$ potentials~\cite{Wi95,St94,Ma96}. When
negative-energy states are included the deuteron binding energy changes
slightly. To compensate for this we adjust the $\sigma$-meson coupling
from the value of the fit in Ref.~\cite{Ma89}, $\frac{g_\sigma^2}{4
  \pi}=8.08$, to $\frac{g_\sigma^2}{4 \pi}=8.55$.

The single-nucleon form factors are taken from the recent work of
Kelly~\cite{Ke02} that includes data from
Jefferson Laboratory on the ratio $G_{E_p}/G_{M_p}$ and $G_{E_n}$.
However, we find that the results are very close to those obtained using the
nucleon form factors of Mergell 
 {\it et al.}~\cite{Me95}. Both parameterizations incorporate 
constraints on the asymptotic
shape of $F_1$ and $F_2$ using arguments from perturbative QCD.

\begin{figure}[h,t,b]
\includegraphics[width=3.2in]{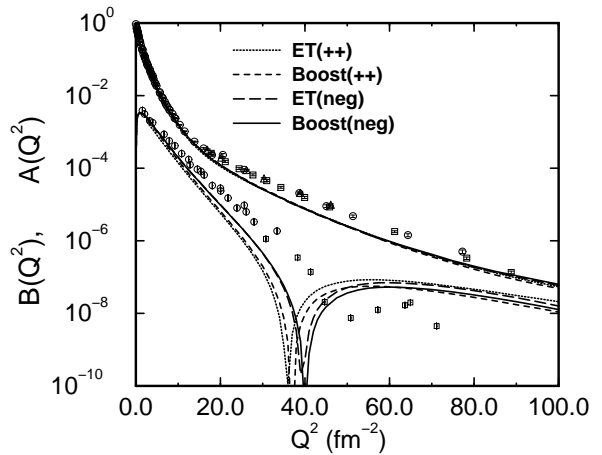}

\caption{The deuteron structure functions $A$ and $B$.  The
  experimental data for $A$ of Refs.~\cite{Adata,AandB} are 
denoted by the open circles, while
  those of Refs.~\cite{Ab00B,Al99,Ab99} are represented by triangles 
and squares.  The experimental data for $B$ of 
Refs.~\cite{AandB,B.expt1.dat} are shown by open circles, while
those of Ref.~\cite{B.expt2.dat} are shown by squares. The dotted
  and short-dashed lines labeled ET(++) and ET(neg) are based on the
  ET calculation without the dynamical boost.  The long-dashed and
  solid curves labeled Boost(++) and Boost(neg) include the dynamical
  boost. 
\label{fig:AandB_Boostfig}  }
\end{figure}

We see clearly in Fig.~\ref{fig:AandB_Boostfig} that the baseline ETIA
calculation underpredicts the $A$ data for $Q^2$ = 25--75 fm$^{-2}$, i.e.
$Q^2=$1--3 GeV$^2$.  This is
true for both the curve labeled ET($++$)---which includes only $++$
states---and the one labeled ET(neg)---which includes the effects of
all positive- and negative-energy states to all orders in $K_{\rm
ET}$. Data from the two JLab experiments that have measured 
$A$~\cite{Al99,Ab99} are denoted by triangles and squares. 
Note that these experiments confirm the trend of the
SLAC data of Arnold {\it et al.}  The extant experimental data for 
$B$~\cite{AandB,B.expt1.dat,B.expt2.dat} 
(which, as yet, include no JLab data) are even
less well-described. Already at $Q^2 \lsim$ 25 fm$^{-2} \approx$ 
1 GeV$^2$ there is significant
disagreement between our ETIA calculation and the data.

In Fig.~\ref{fig:AandB_Boostfig} we also present results for
observables $A$ and $B$ showing the effects of the approximate dynamical boost.
The curves are labeled Boost($++$) and Boost(neg).  The net effect of
the dynamical boost is significant but not large.  For either the $++$
calculation or the one including all states, the boost shifts the
minimum in the calculated $B$ observable to somewhat higher $Q^2$.  As seen in
Fig.~\ref{fig:AandB_Boostfig}, the $A$ observable is not much altered
by the boost.  These results suggest that further refinement of the
approximate boost developed here, while theoretically interesting, would
have little phenomenological impact on these observables.

\begin{figure}[h,t,b]
\includegraphics[width=3.2in]{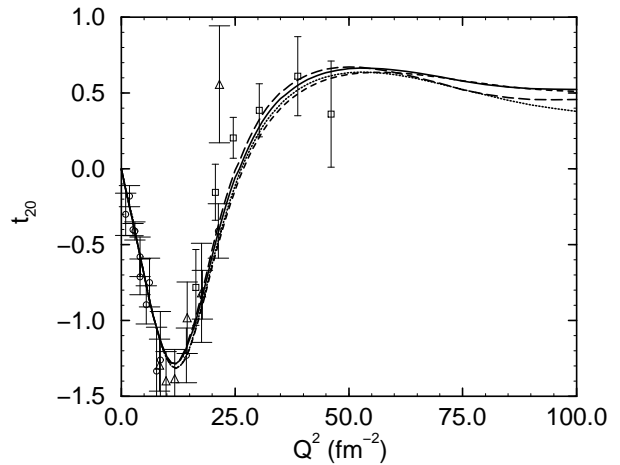}

\caption{The tensor-polarization observable $t_{20}$. The older
  experimental data~\protect{\cite{T20.expt1.dat}} and the NIKHEF data of 
  Bouwhuis {\it et al.}~\protect{\cite{Bo95}} are shown by
circles, the JLab data
  of Abbott {\it et al.}~\protect{\cite{Ab00A}} by squares and the 
Novosibirsk data of Nikolenko {\it et al.}~\protect{\cite{Ni03}} 
by triangles. The lines have the
  same meaning as in Fig.~\ref{fig:AandB_Boostfig}. 
\label{fig:T20_Boostfig}   }
\end{figure}

A similar result is seen for observable $t_{20}$ in
Fig.~\ref{fig:T20_Boostfig}.  Although effects on observables are
not large, inclusion of the approximate dynamical boost in our calculations is an
important step forward because it ensures that the deuteron mass is 
independent of frame, whereas    
an {\it ad hoc} factor $\lambda({\bf P}^2)$ in $K_{\rm ET}$ was needed
for this purpose in prior work.

\begin{figure}[h,t,b]
\includegraphics[width=3.2in]{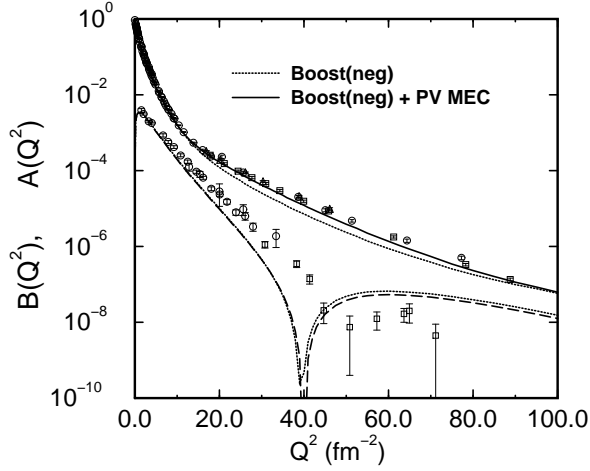}
\caption{The deuteron structure functions $A$ and $B$ 
with and without the PV-coupling MEC. Experimental data points
are as described in caption to 
Figure~\ref{fig:AandB_Boostfig}. 
\label{fig:AandB_PVpiMECfig}  }
\end{figure}

\begin{figure}[h,t,b]
\includegraphics[width=3.2in]{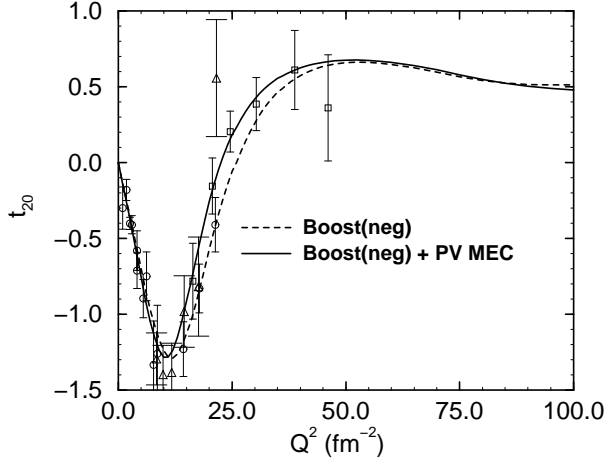}

\caption{The tensor-polarization observable $t_{20}$
with and without the PV-coupling MEC. Both calculations include all
positive- and negative-energy sectors.  Experimental data points are as
described in caption of Figure~\ref{fig:T20_Boostfig}. 
\label{fig:T20_PVpiMECfig}  }
\end{figure}

Calculations with and without the PV-coupling MEC are shown in
Fig.~\ref{fig:AandB_PVpiMECfig} for $A$ and $B$ and in
Fig.~\ref{fig:T20_PVpiMECfig} for $t_{20}$.  This MEC, which must be
included because we chose to employ pseudovector $\pi N$ coupling,
provides a significant increase in the $A$ observable, small changes
in $B$ and a significant shift of $t_{20}$ toward smaller $Q$.  The
increase in $A$ helps to move the result into better agreement with
the experimental data. But our results for $B$ still decrease too fast
with $Q$ and the minimum occurs at too low a value of $Q$.  The shift
of $t_{20}$ aligns our results quite well with the JLab data of Abbott
{\it et al.}~\cite{Ab00A,Ab00B}.

\begin{figure}[h,t,b]
\includegraphics[width=3.2in]{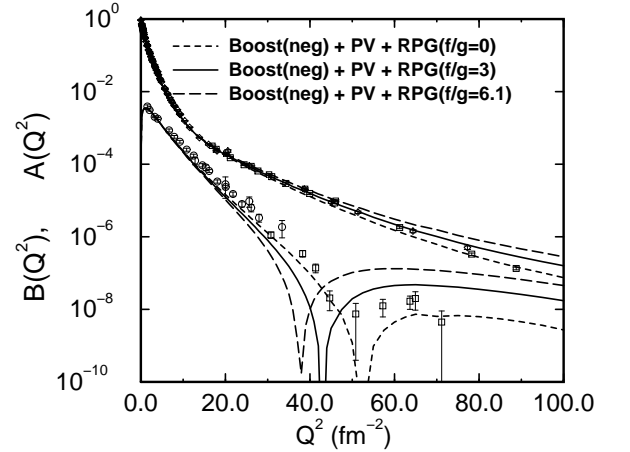}
\caption{The deuteron structure functions $A$ and $B$.
Each line is based on including all positive and negative-energy states,
the boost, 
the PV-coupling MEC and the $\rho\pi\gamma$ MEC.  Lines
differ only by the value of tensor $\rho N$ coupling $f_\rho/g_\rho$ used in
the $\rho\pi\gamma$ graph.  
\label{fig:AandB_neg_PV_RPG=0to6}  }
\end{figure}

\begin{figure}[h,t,b]
\includegraphics[width=3.2in]{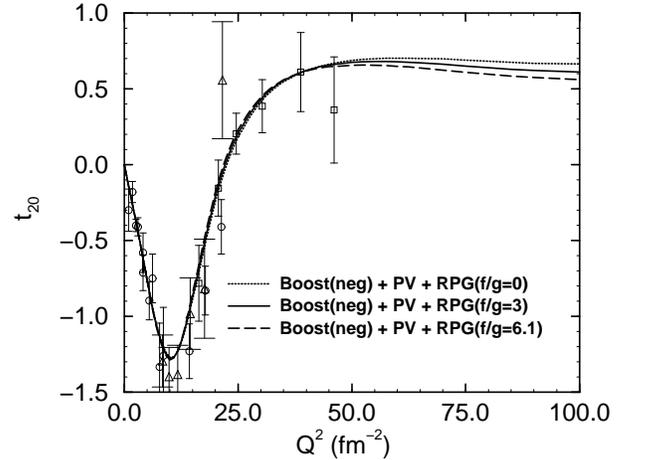}
\caption{The tensor-polarization observable $t_{20}$.
Each line is based on including all positive and negative-energy states,
the boost, 
the PV-coupling MEC and the $\rho\pi\gamma$ MEC.  Lines
differ only by the value of tensor $\rho N$ coupling $f/g$ used in
the $\rho\pi\gamma$ graph.  
\label{fig:T20_neg_PV_RPG=0to6} }
\end{figure}
In Fig.~\ref{fig:AandB_neg_PV_RPG=0to6} we present our results for $A$
and $B$ when the $\rho \pi \gamma$ MEC is included using the
vector-meson-dominance form factor (\ref{eq:rpgf}) as in the work of
Hummel and Tjon.  Results are shown for three values of the tensor
coupling ratio $f_{\rho}/g_{\rho} =$ 0, 3, and 6.1 and all
calculations include the full set of positive- and negative-energy
states and the PV-coupling MEC.  At larger values of $Q$, the
calculated values of $A$ are increased by the $\rho \pi \gamma$ MEC.
For the ratio $f/g = 0$, the $\rho\pi\gamma$ contribution provides
good agreement with the data for observable $A$.

Meanwhile, Fig.~\ref{fig:T20_neg_PV_RPG=0to6} shows results for the tensor
observable $t_{20}$, which is rather insensitive to the
$\rho\pi\gamma$ contribution.  In contrast, this exchange current has
a significant impact on the observable $B$. In particular, different
choices for the tensor $\rho N$ coupling change the results markedly,
as is evident in the lower curves of
Fig.~\ref{fig:AandB_neg_PV_RPG=0to6}.  It is clear from these results
that the minimum of $B$ is especially sensitive to the value of
$f_\rho/g_\rho$, and that experimental data are described best with
zero tensor coupling.  However, a softer $\rho\pi\gamma$ form factor
than that of Eq.~(\ref{eq:rpgf}) would reduce the impact of the $\rho
\pi \gamma$ MEC on all observables. Also, Hummel and Tjon~\cite{HT}
have estimated the effects of a possible $\omega\sigma\gamma$
meson-exchange current and found that it can have a significant effect
on the minimum of the $B$ observable when a vector-meson-dominance
$\omega \sigma \gamma$ form factor is used.  Clearly, a better
understanding of the effects of meson-exchange corrections on the
electron-deuteron $B$ structure function is needed.  Lattice QCD
calculations along the lines of Ref.~\cite{Ed04} are needed for the
$\rho\pi\gamma$ form factor with $Q^2$ values up to 2 GeV$^2$. 

Negative-energy states have significant effects on $A$ and
(particularly) $B$. See, e.g. the curves labeled Boost($++$) and
Boost(neg) in Fig.~\ref{fig:AandB_Boostfig}.  We do not, however, find
as large an effect from negative-energy states as was obtained in the
work of van Orden, Devine and Gross~\cite{vO95}. There the inclusion
of such effects produced an acceptable description of the $B$ data.
This is likely due in part to the fact that we use a pseudovector $\pi
N$ coupling, whereas in Ref.~\cite{vO95} an admixture of about 25\%
pseudoscalar $\pi N$ coupling was employed.  Pseudoscalar $\pi N$
coupling is known to produce larger effects from negative-energy
states.

\begin{figure}[h,t,b]
\includegraphics[width=3.2in]{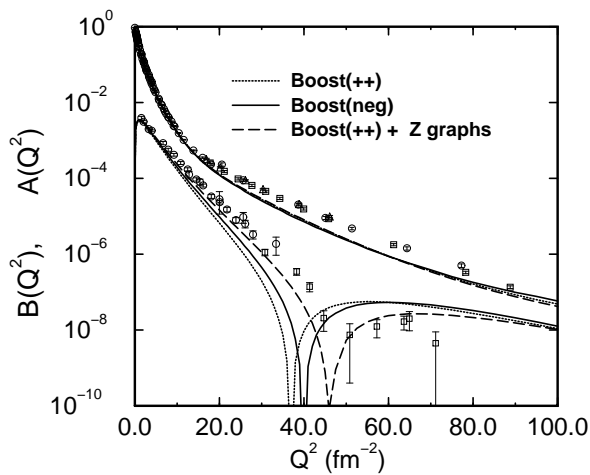}
\caption{The deuteron structure functions $A$ and $B$.
In all cases the boost is included and the meson-exchange corrections are omitted.  
Dotted lines are
based on including $++$ states only, solid lines are based
on including all positive and negative-energy states and dashed
lines are based on including $++$ states and adding 
the Z graphs in first-order perturbation 
theory as in Eq.~(\ref{eq:Zgraphs}). 
\label{fig:AandB_Boost++_Z_neg}  }
\end{figure}

\begin{figure}[h,t,b]
\includegraphics[width=3.2in]{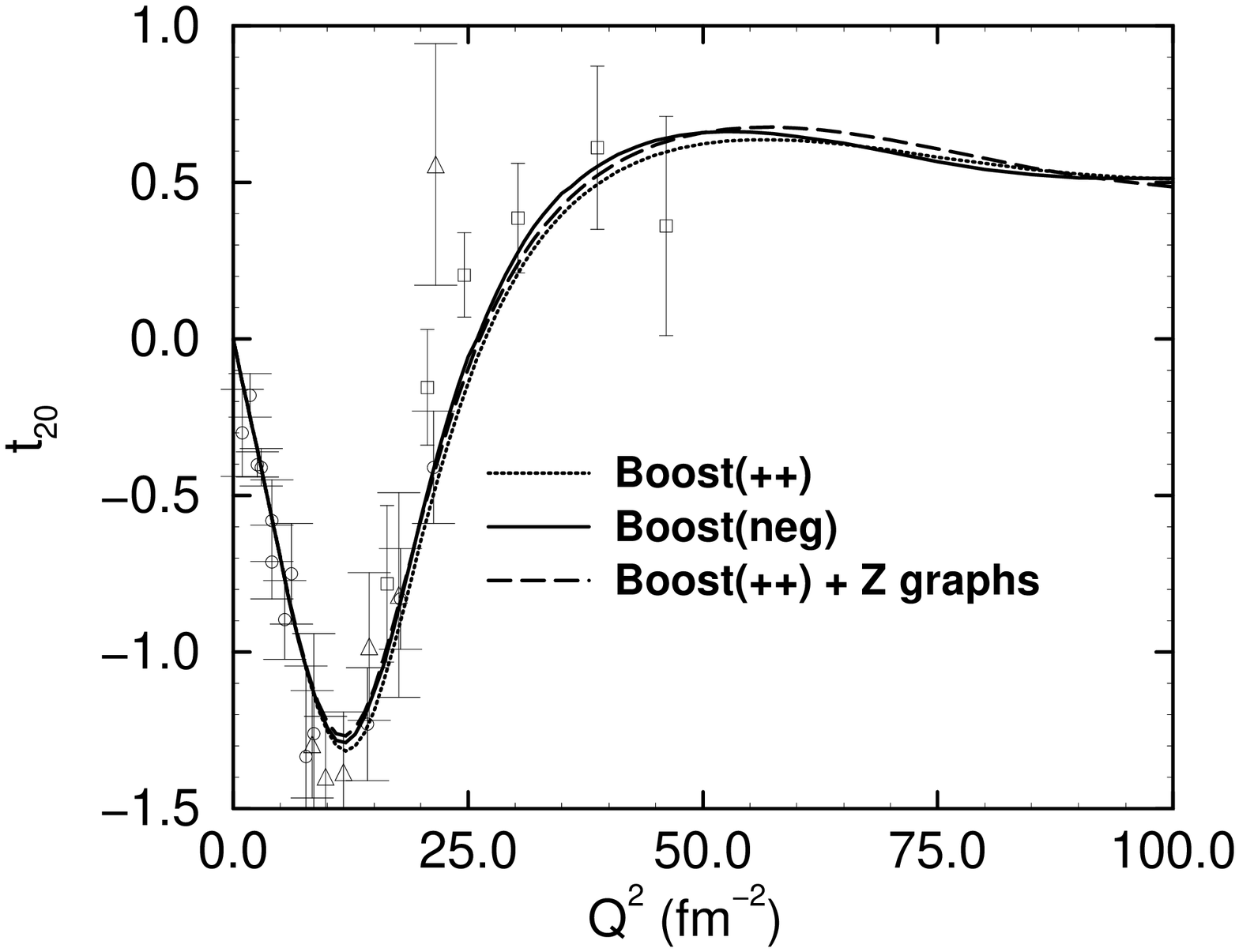}
\caption{The tensor-polarization observable $t_{20}$.
Lines have the same meaning as in Fig~\ref{fig:AandB_Boost++_Z_neg}.
\label{fig:T20_Boost++_Z_neg}  }
\end{figure}

In order to clarify the significance of including all positive and
negative-energy states in our analysis, we also calculate the
leading-order Z-graphs.  Figs.~\ref{fig:AandB_Boost++_Z_neg} and
\ref{fig:T20_Boost++_Z_neg} compare results obtained using $++$ states
only, adding the leading-order Z-graphs as in Eq.~(\ref{eq:Zgraphs}),
and including all positive and negative-energy states.  Meson-exchange
currents are omitted.  In the case of observable $B$ the leading-order
Z-graphs do not accurately reproduce the results of a calculation
which includes all positive- and negative-energy states to all orders
in perturbation theory.  First-order perturbation theory for Z-graphs
is adequate for observables $A$ and $t_{20}$, but not for $B$.
Solving for the deuteron vertex functions non-perturbatively in all
positive- and negative-energy sectors one finds smaller effects on the
$B$ observable due to couplings to negative-energy states than are
predicted by leading-order Z-graphs.

Our results suggest that the existing data for the deuteron's $A$ and
$t_{20}$ observables are described reasonably well when relativistic
dynamics, boost effects and the PV-coupling current are implemented in
a consistent formalism and calculation.  However, the data for the $B$
observable are not explained unless there is some non-standard
contribution, such as the $\rho\pi\gamma$ MEC with $f_\rho=0$.  Other
analyses have suggested that various relativistic effects, effects of
negative-energy states or effects of the PV-coupling current might be
larger than we find, but we believe our examination of each of these
effects to be reliable within the ET formalism.

If a vector-meson-dominance $\rho \pi \gamma$ form factor is correct,
then the $\rho \pi \gamma$ MEC can help to explain the $B$ data
provided that a very small tensor $\rho N$ coupling is used.  The
ratio $f_{\rho}/g_{\rho}$ should be substantially less than the 6.1
used in one-boson-exchange models of the $NN$ interaction, or even a
value which would explain the isovector magnetic moment of the
nucleon.  Fig.~\ref{fig:T20_neg_PV_RPG=0to6} shows that the $t_{20}$
data of Refs.~\cite{Ab00A,T20.expt1.dat,Bo95,Ni03} are well-described by our
approach out to $Q^2 \approx 2 \, {\rm GeV}^2$.  This observable is
fairly insensitive to some of the dynamics that plays a role in $A$
and $B$ (single-nucleon form factors, the $\rho \pi \gamma$
MEC). However, it is quite sensitive to relativistic effects (see, for
instance~\cite{Wi95,CK99}), so it is gratifying that our approach
reproduces the data, especially that of Ref.~\cite{Ab00A} at large
$Q^2$, so well.

\acknowledgments{We thank J. Adam, E.~J.~Beise, R.~Edwards,
  M.~Petratos, and J.~A.~Tjon for useful
  conversations. D.~R.~P. thanks both the Theory Group at Jefferson
  Lab and the TQHN group at the University of Maryland for their
  hospitality during this work. We are also grateful to the
  U.~S. Department of Energy, Nuclear Physics Division, for its
  support under grants DE-FG02-93ER-40756, DE-FG02-02ER41218,
  DE-FG02-93ER-40762, and DE-FG03-97ER41014.}


\begin{thebibliography}{10}
\bibitem{m4} Strict expansions in
powers of $1/m$ should be avoided because corrections like 
$p^4/m^4$ are not bounded as $p \rightarrow \infty$.
Our expansions generally involve bounded operators. We refer to
typical bounded correction terms, such as $p^4/\epsilon^4$, 
as order $1/m^4$ for simplicity.
\end{thebibliography}
\end{document}